\documentclass[11pt]{article}
\pdfoutput=1 
\usepackage{jheppub}
\usepackage{bm,mathtools, slashed}
\usepackage{physics}
\usepackage{siunitx}

\newcommand{\diag}{\mathop{\mathrm{diag}}}
\newcommand{\bB}{\bm{B}}
\newcommand{\bk}{\bm{k}}
\newcommand{\bp}{{\bm{p}}}
\newcommand{\LepTdensity}{h}
\newcommand{\tilLepTdensity}{\tilde{h}}
\newcommand{\LepT}{H}
\newcommand{\tilLepT}{\tilde{H}}
\newcommand{\Ffunc}{{\mathcal{A}}}
\newcommand{\Gfunc}{{\mathcal{B}}}
\newcommand{\Hfunc}{{\mathcal{C}}}
\newcommand{\Kfunc}{{\mathcal{D}}}

\newcommand{\sgn}{\mathop{\mathrm{sgn}}}
\newcommand{\Dn}{S}
\newcommand{\ri}{\mathrm{i}}
\newcommand{\re}{\mathrm{e}}

\title{Photon polarization tensor at finite temperature and density in a magnetic field}

\preprint{YITP-24-153, RIKEN-iTHEMS-Report-24}

\author[a]{Kenji Fukushima}

\author[b,c]{Yoshimasa Hidaka}

\author[a]{Tomoya Uji}

\affiliation[a]{Department of Physics, The University of Tokyo, 7-3-1 Hongo, Bunkyo-ku, Tokyo 113-0033, Japan}

\affiliation[b]{Yukawa Institute for Theoretical Physics, Kyoto University, Kyoto 606-8502, Japan}
\affiliation[c]{Interdisciplinary Theoretical and Mathematical Sciences Program (iTHEMS), RIKEN, Wako, Saitama 351-0198, Japan}

\emailAdd{fuku@nt.phys.s.u-tokyo.ac.jp}
\emailAdd{yoshimasa.hidaka@yukawa.kyoto-u.ac.jp}
\emailAdd{uji@nt.phys.s.u-tokyo.ac.jp}

\abstract{
We present analytical and numerical calculations for the photon polarization tensor at finite temperature and density in a constant magnetic field.
We first discuss the tensor decomposition in the presence of the magnetic field, which breaks rotational symmetry.
Then, we analytically perform all the momentum integrations and numerically take the Landau level sum.
We confirm that the imaginary part of the photon polarization tensor correctly reproduces the known result from the independent calculation. 
We utilize the Kramers-Kronig relation to estimate the real part numerically as a function of the momenta, the chemical potential, and the finite temperature.
As an application, we consider the real photon limit and estimate the photon decay rate and the Stokes parameter in the hot and dense medium.
We specifically quantify the difference between the X-mode and the O-mode with the polarization orthogonal and parallel to the magnetic field.  As long as the magnetic field is weak, the decay rate of the X-mode photon is larger than that of the O-mode photon, while the O-mode becomes dominant due to the Landau level suppression of the X-mode at a strong magnetic field.
We also find that the eigenmodes of the propagating photon change their polarization state with increasing density.
}

\begin{document}
\maketitle

%%%%%%%%%%
\section{Introduction}

% general introduction
% magnetic catalysis
% phase diagram
%   Fraga, Ferrer-Incera, Ruggieri, Pawlowski
% inverse magnetic catalysis (lattice QCD)
%   Bali, Endrodi, A. Schoefer
% hadron spectra in LQCD with B
%   H.-T. Ding
Both theoretically and phenomenologically, magnetic fields are important probes across a wide range of fields in physics, including condensed matter physics, astrophysics, and nuclear physics.
Physics in quantum chromodynamics (QCD) is rich and complicated, and various aspects of QCD have been revealed with magnetic field probes.
To give some examples, theoretical investigations on the QCD phase diagram in the presence of an external magnetic field $B$ have been addressed~\cite{Fraga:2008qn,Mizher:2010zb,Ferrer:2013noa,Gatto:2010qs,Gatto:2010pt,Ferreira:2014kpa,Mueller:2015fka}, and enhancement of chiral symmetry breaking by magnetic fields is known as the magnetic catalysis and has been discussed over decades~\cite{Klimenko:1991he,Gusynin:1994re,Gusynin:1995nb,Shushpanov:1997sf,Fukushima:2012xw,Bali:2013txa}; see ref.~\cite{Miransky:2015ava} for a review.
Besides, the Monte-Carlo simulation in QCD with external magnetic fields has been conducted without the sign problem~\cite{DElia:2010abb,Bali:2011qj}; see ref.~\cite{Endrodi:2024cqn} for the state-of-the-art overview.
The simulation has provided us with deep insights into QCD phenomena induced by the magnetic fields; the inverse magnetic catalysis was discovered in lattice QCD at high temperature~\cite{Bali:2011qj,Bali:2012zg,Bruckmann:2013oba}, the meson mass spectra under magnetic background invoked inspiring discussions on the physical interpretation~\cite{Hidaka:2012mz,Hattori:2015aki,Ding:2020jui},
and so on.

% heavy-ion (hot plasma)
%   Skokov + ?, X.-G. Huang + ?, McLerran-Skokov
% STAR
% D meson flow --> B (STAR)
% spin polarization (Lambda - Lambda-bar) (STAR)
% Ultra Peripheral Collision (UPC) Brandenburg, ...
% Anomalous transport (CME, CSE, ...)
%   J. Liao
The phenomenological aspects of magnetic fields have attracted a lot of attention, especially in the theoretical community, in demand from real experiments.
Indeed, in non-central heavy-ion collisions at Relativistic Heavy Ion Collider (RHIC) and Large Hadron Collider (LHC), strong magnetic fields $\sim10^{18}$\,G (comparable to $m_\pi^2$) should be realized~\cite{Skokov:2009qp,Voronyuk:2011jd,Deng:2012pc,McLerran:2013hla}.
Since the energy scales of $B$ and the temperature $T$ are larger than the light-flavor quark mass, a quark-gluon plasma (QGP) can be regarded as a chiral material that exhibits nontrivial phenomena.
For example, several anomalous transport effects as represented by the chiral magnetic effect~\cite{Kharzeev:2007jp,Fukushima:2008xe}, the chiral separation effect~\cite{Son:2004tq,Metlitski:2005pr}, and the chiral vortical effect~\cite{Son:2009tf,Kharzeev:2010gr} are believed to emerge from properties of the chiral materials~\cite{Kharzeev:2004ey,Kharzeev:2007jp,Kharzeev:2013ffa,Shi:2017cpu}; see ref.~\cite{Kharzeev:2015znc} for a review.
Importantly, these effects are attributed to macroscopic manifestations of the chiral anomaly and QCD topological structures.
It should be noted that the large vorticity, an analogue of the magnetic field, has been confirmed experimentally by hyperon global polarization measured by the STAR collaboration at RHIC~\cite{STAR:2017ckg} as predicted by the theoretical work~\cite{Liang:2004ph}.
The estimate of vorticity assumes a formula derived in ref.~\cite{Becattini:2016gvu} in which the magnetic effect of the intrinsic magnetic moments of created particles was also discussed.

% cores in the neutron star with B
%   Ferrer-Incera, C. Manuel, Shovkovy-Noronha
%   Fukushima-Warringa
% B origin
%   Ohnishi-Yamamoto / Reddy + ?
% CSC with B
%   Alford, Blaschke
% Brauner-Yamamoto (high density + B --> CSL)
% --> IXPE

It would be of great significance to investigate magnetized QCD matter not only along the temperature axis. 
Cold and dense quark matter, such as a color superconductor with strong magnetic fields, might be realized in cores of neutron stars~\cite{Alford:1999pb,Ferrer:2005vd,Ferrer:2006vw,Noronha:2007wg,Fukushima:2007fc}.
The strength of magnetic fields at the surface of specific neutron stars called magnetars is estimated to be of order $10^{15}$\,G\@.
Since the magnetic flux in the central cores of the stars is squeezed, the field strength is thought to be as high as $\sim10^{18}$\,G\@.
It is conceivable that such strong $B$ affects the ground state properties in QCD matter.
One plausible candidate for the true ground state is the chiral soliton lattice, which turns out to be energetically favored at high density and strong magnetic field~\cite{Brauner:2016pko,Chen:2021vou,Qiu:2023guy,Evans:2023hms}.

% new section about IXPE
% more details -- what is necessary for estimating the polarization
In addition, the magnetic fields on magnetar surfaces have drawn increasing attention in recent years.  This is caused by the possibility that we can estimate the strength and geometry of the magnetars' magnetic fields through the X-ray polarization observation from the Imaging X-ray Polarimetry Explorer (IXPE)~\cite{Taverna:2020vpr,Taverna:2022jgl,Taverna:2024uop}.
In general, photons under strong magnetic fields have two distinct modes with linear polarizations; namely, the ordinary mode (O-mode) with polarization parallel to the magnetic field and the extraordinary mode (X-mode) with polarization perpendicular to the magnetic field.
From the observational data from IXPE, the polarization angle and polarization degree of X-ray have been extracted as functions of the energy of photons.
With this extracted information, we could reconstruct the profile of the magnetic fields of magnetars.
Indeed, some models are proposed to explain the observational data.
For example, in the resonance Compton scattering scenario~\cite{Taverna:2022jgl}, the cross section of the Compton scattering between photons and electrons becomes large when the energy of photons is comparable to the cyclotron frequency of electrons.
As a result, the resonant cross section is mode-sensitive, and this scenario yields the polarization degree $\sim 33\%$ of X-rays from the magnetosphere.
However, according to the IXPE results~\cite{Taverna:2022jgl,Zane:2023khc,Turolla:2023ruu}, the polarization degree reaches as large as $80\%$, and the energy dependence of the polarization degree was found to be sizable.
It is further puzzling that the polarization angle swings by $\ang{90}$ as the energy of photons changes.
To make a consistent view, the mode conversion scenario was discussed in refs.~\cite{Lai:2003nd,Lai:2022knd}.
In this scenario, the combination of the vacuum birefringence effects and the plasma birefringence effects results in a stochastic conversion of the O-mode into the X-mode of photons, and vice versa.
Although this scenario can successfully explain the observation qualitatively, it is desirable to improve it to make a quantitative match with the data.
In order to supplement this model with electron contribution corrections, the field-theoretical evaluation of the polarization tensor with electron loops is necessary.
To this end, we calculate the polarization tensor in the framework of quantum field theory at finite temperature and density.

% Polarization in B
% Ferrer-Incera / Hattori-Itakura / Russian / Hidaka / Shovkovy

As mentioned above, there are so many physical systems where the magnetic fields play an important role, which has motivated a lot of theoretical work involving magnetic fields.
In particular, the field-theoretical calculation in the presence of external electromagnetic fields was founded by Schwinger, which can be traced back to the old days of Euler and Heisenberg.
They suggested that the QED vacuum filled with electrons and positrons is modified by the polarization with strong electromagnetic fields~\cite{Heisenberg:1936nmg,Schwinger:1951nm,Dunne:2004nc,Ferrer:2012pb}.
More recently, the vacuum contribution of the polarization tensor in magnetic fields was intensively discussed in refs.~\cite{Baier:2007dw,Hattori:2012je,Hattori:2012ny}.
For the finite temperature, $T\neq0$, and zero chemical potential, $\mu=0$, the strong field approximation for the polarization tensor was argued in ref.~\cite{Alexandre:2000jc}, and the full Landau level calculation for the imaginary part of the polarization tensor is found in refs.~\cite{Wang:2020dsr,Wang:2021ebh}, which has been extended to finite density, $\mu\neq 0$~\cite{Wang:2021eud}.
The calculation would become simple in the lowest Landau level approximation as argued in refs.~\cite{Hattori:2022uzp,Hattori:2022wao}.
In addition, for $T\neq 0$ and $\mu \neq 0$, the dynamical polarization function of graphene in a magnetic field was treated in ref.~\cite{Pyatkovskiy:2010xz}.
The relation between the eigenmodes of propagating photons and the polarization tensor in a magnetic field for $T\neq 0$ and $\mu \neq 0$ was discussed in~\cite{PerezRojas:1979jrk}.
The present work aims to complete the full analysis of the real part of the polarization tensor as well as the imaginary part at finite $T$ and $\mu$, taking account of all Landau levels.

%%%%%%%%%%
\section{Formulation}

This paper aims to complete the full evaluation of the photon polarization tensor with electrons and positrons at finite temperature and
density.  We will proceed to the technical details of the explicit derivation in the next section.  Here, we shall summarize basic formulas that would be convenient for clarifying our convention.

%%%%%%%%
\subsection{Propagator in the magnetic field}

The electron field, $\psi(x)$, satisfies the following Dirac equation:
\begin{equation}
    (\ri\slashed{D}-m)\psi(x)=0\,,
\end{equation}
where $m$ is the electron mass and $\slashed{D}\coloneqq\gamma^\mu D_\mu$ is Feynman's slash.  The Dirac matrices, $\gamma^{\mu}$, satisfy the Clifford algebra, $\{\gamma^\mu,\gamma^\nu\}=2\eta^{\mu\nu}$, where $\eta_{\mu\nu}=\diag(1,-1,-1,-1)$ in our convention.  The covariant derivative for the electron is $D_\mu\coloneqq\partial_\mu -\ri e A_\mu$ with $e>0$, i.e., the electron charge is $-e$.  We choose the $z$-axis along the external magnetic field direction; $\bB=(0,0,B)$ with $B>0$.  Throughout this paper, we will work in a symmetric gauge, $A^\mu =(0,-yB/2,xB/2,0)$.  In our notation, we write the four-momentum simply as $p^\mu=(p_0,p_x,p_y,p_z)$ using the coordinate subscripts and define the transverse and the longitudinal four momenta as $p^\mu_\perp=(0,p_x,p_y,0)$ and $p^\mu_\parallel=(p_0,0,0,p_z)$, respectively.  We note that $p_\perp^2=-p_x^2-p_y^2=-\bp_\perp^2$ and $p_\parallel^2=p_0^2-p_z^2$.  In the imaginary-time formalism of quantum field theory at finite temperature $T$ and chemical potential $\mu$, we may replace $p^\mu\to \tilde{p}^\mu\coloneqq (\ri\nu_n+\mu,\bp)$ with the fermionic Matsubara frequency given by $\nu_n=(2n+1)\pi T$ ($n\in\mathbb{Z}$).
Then, the electron propagator in the imaginary-time formalism reads:
\begin{equation}
    S(\tilde{p}) = \sum_{l=0}^\infty \frac{-\Dn_{l}(\tilde{p})}{\tilde{p}_\parallel^2- m_l^2}\,.
    \label{eq:propagator}
\end{equation}
We note that the Schwinger phase is dropped because it is irrelevant in calculating the polarization tensor.  In the denominator, $m_l^2\coloneqq 2eBl + m^2$ is the effective transverse mass at the Landau level $l$.  The numerator on the right-hand side is
\begin{equation}
  \Dn_{l}(p) \coloneqq (\slashed{p}_\parallel+m)\bigl[ P_- A_{l-}(\xi_p) + P_+ A_{l+}(\xi_p) \bigr] + \slashed{p}_\perp B_l(\xi_p)\,,
  \label{eq:Dn}
\end{equation}
where we define the following functions:
\begin{subequations}
\begin{align}
 A_{l-}(\xi_p) &\coloneqq 2\re^{-2\xi_p}(-1)^l L_{l}(4\xi_p)\,,\\
 A_{l+}(\xi_p) &\coloneqq 2\re^{-2\xi_p}(-1)^{l-1} L_{l-1}(4\xi_p)\,,\\
 B_{l}(\xi_p) &\coloneqq 4\re^{-2\xi_p}(-1)^{l-1} L_{l-1}^{(1)}(4\xi_p)\,,
\end{align}
\end{subequations}
where a dimensionless variable, $\xi_p \coloneqq \bm{p}^2_\perp/(2eB)$, is introduced for notational brevity.
In this work, we adopt the following definition for the generalized Laguerre polynomials:
\begin{equation}
    L_n^{(\alpha)}(x) \coloneqq \frac{\re^x x^{-\alpha}}{n!}\dv[n]{x}\re^{-x} x^{n+\alpha}
    \label{eq:LAlphanN}
\end{equation}
and the Laguerre polynomials are $L_n(x) = L_n^{(0)}(x)$ in particular.
In our present convention, $L_{-n}^{(\alpha)}(x)=0$ for $n>0$ should be understood.  It is thus evident from eq.~\eqref{eq:Dn} that the Landau zero mode exists only for one spin state projected by $P_-$. 
The quantity inside the square brackets contains the projection operators of the spin along the $z$-axis (for $eB>0$):
\begin{equation}
    P_{\pm} \coloneqq \frac{1}{2} \bigl( 1\pm \ri\gamma^1\gamma^2 \bigr)\,.
\end{equation}

%--- figure ---%
\begin{figure}
    \centering
    \includegraphics[width=0.5\textwidth]{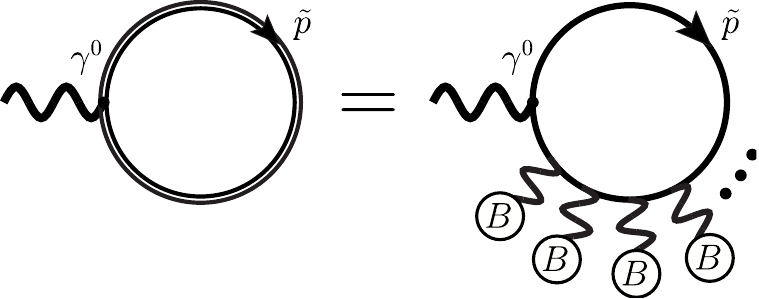}
    \caption{Tadpole diagram for calculating the expectation value of the density.}
    \label{fig:tadpole}
\end{figure}
%--- figure ---%

Before dealing with the polarization tensor, which has 16 components, let us consider the calculation of the density, $\langle n(x)\rangle=\langle\bar{\psi}(x)\gamma^0\psi(x)\rangle$, as a one-component warm-up exercise.  Figure~\ref{fig:tadpole} shows the diagram corresponding to $\langle n(x)\rangle$.  We use the double line to represent the fermion propagator in the presence of background $\bB$ as schematically illustrated in figure~\ref{fig:tadpole}.  Due to translational symmetry, we can take $x=0$ to simplify the expression slightly as
\begin{equation}
  \langle n(0)\rangle = (-1) T\!\! \sum_{n=-\infty}^\infty \sum_{l=0}^\infty
  \int\frac{\dd[3]{p}}{(2\pi)^3}\frac{-\tr\bigl[ \gamma^0 S_l(\tilde{p})\bigr]}{\tilde{p}_\parallel^2-m_l^2}\,.
\end{equation}
The overall minus sign appears from the fermion loop.  Then, it is straightforward to take the trace using eq.~\eqref{eq:Dn} to find
\begin{align}
  \langle n(0)\rangle
  &= \sum_{l=0}^\infty \int\frac{\dd[2]{p_\perp}}{(2\pi)^2}\, \re^{-2\xi_p}(-1)^l \Bigl[ L_l(4\xi_p) - L_{l-1}(4\xi_p) \Bigr] T\!\! \sum_{n=-\infty}^\infty
  \int\frac{\dd{p_z}}{2\pi} \frac{4\tilde{p}^0}{\tilde{p}_\parallel^2 - m_l^2} \notag\\
  &= \frac{eB}{8\pi}\sum_{l=0}^\infty \int_0^\infty \dd{u}\, \re^{-u/2}(-1)^l \Bigl[ L_l(u) - L_{l-1}(u) \Bigr] \times\notag\\
  &\qquad\qquad\qquad\qquad \times 
  \int_{-\infty}^\infty\frac{\dd{p_z}}{2\pi}\, 2\Bigl[ n_\mathrm{F}(E_l(p_z)-\mu)-n_\mathrm{F}(E_l(p_z)+\mu)\Bigr] \notag\\
  &= \frac{eB}{2\pi} \sum_{l=0}^\infty \alpha_l \int_{-\infty}^\infty \frac{\dd{p_z}}{2\pi}\Bigl[n_\mathrm{F}(E_l(p_z)-\mu)-n_\mathrm{F}(E_l(p_z)+\mu)\Bigr]\,,
  \label{eq:numberdensity}
\end{align}
where $E_l(p_z)\coloneqq\sqrt{p_z^2+m_l^2}$ is the electron energy at the Landau level $l$.  From the first to the second line, $u=4\xi_p$ is introduced, so that we can use the formula: $\int_0^\infty \dd{u}\, \re^{-u/2} L_l(u)=2(-1)^l$.  The Matsubara sum leads to $n_\mathrm{F}$ that is the Fermi-Dirac distribution function:
\begin{equation}
  n_\mathrm{F}(E) \coloneqq \frac{1}{\re^{E/T}+1}\,.
\end{equation}
In the last line $\alpha_l$ represents the spin degeneracy factor, i.e., $\alpha_0=1$ and $\alpha_{l>0}=2$.  We can easily check that the above expression for $\langle n(0)\rangle$ is reduced to the standard expression in the limit of vanishing $B$.

%--- figure ---%
\begin{figure}
    \centering
    \includegraphics[width=0.28\textwidth]{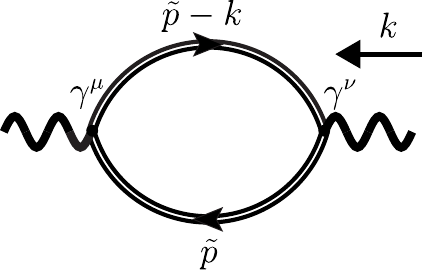}
    \caption{One-loop diagram for the photon polarization tensor.}
    \label{fig:polarization}
\end{figure}
%--- figure ---%

%%%%%
\subsection{One-loop polarization tensor}

It is an immediate extension to apply the above technique to the polarization diagram at the one-loop level as depicted in figure~\ref{fig:polarization}.  For the external momentum, $k^\mu=(\ri\omega_n, \bk)$ with $\omega_n=2n\pi T$ the bosonic Matsubara frequency, the polarization tensor reads:
\begin{equation}
  \Pi^{\mu\nu}(\ri\omega_n,\bk) = (-1) e^2T\sum_{m=-\infty}^\infty \sum_{k,l=0}^\infty\int \frac{\dd[3]{p}}{(2\pi)^3}
  \frac{\tr\bigl[\gamma^\mu \Dn_{k}(\tilde{p})\gamma^\nu \Dn_{l}(\tilde{p}-{k})\bigr]}{(\tilde{p}_\parallel^2-m_{k}^2)[(\tilde{p}_\parallel-{k}_\parallel)^2-m_{l}^2]}\,.
\end{equation}
Here again, the overall $(-1)$ appears from the fermion loop.  
We note $\tilde{p}_0=\ri\nu_m+\mu$, where $\nu_m = (2 m +1)\pi$ is the fermionic Matsubara frequency (not to be confused with the mass $m$). 
For notational simplicity, we represent the numerator by the following symbol,
\begin{equation}
    \LepTdensity_{kl}^{\mu\nu}(p,k) \coloneqq \tr[\gamma^\mu \Dn_{k}(p)\gamma^\nu \Dn_{l}(p-k)]\,.
    \label{eq:lepton_tensor}
\end{equation}
 Since the denominator does not depend on transverse $p_\perp^\mu$, we can factorize the momentum integral as
\begin{equation}
  \Pi^{\mu\nu}(\ri\omega_n,\bm{k})
  = (-1) e^2T\sum_{m=-\infty}^\infty \sum_{k,l=0}^\infty \int \frac{\dd{p_z}}{2\pi} \frac{\LepT_{kl}^{\mu\nu}(\tilde{p}_0,p_z;k_\parallel,z)}{(\tilde{p}_\parallel^2-m_{k}^2)[(\tilde{p}_\parallel-{k}_\parallel)^2-m_{l}^2]}\,,
\end{equation}
where we define the numerator, i.e., the lepton tensor integral, as
\begin{equation}
\label{eq:h-kl}
  \LepT_{kl}^{\mu\nu}(\tilde{p}_0,p_z;k_\parallel,z) \coloneqq \int\frac{\dd[2]{p_\perp}}{(2\pi)^2}\, \LepTdensity_{kl}^{\mu\nu}(\tilde{p},k)\,,
\end{equation}
where $z\coloneqq\bm{k}_\perp^2/(2eB)$.
%It should be noted that $z\coloneqq\bm{k}_\perp^2/(2eB)$ in the above, which is different from $z$ in eq.~\eqref{eq:numberdensity}.

In later discussions, we shall evaluate $\LepT_{kl}^{\mu\nu}(\tilde{p}_0,p_z; k_\parallel,z)$, which is the central part in this paper, and for the moment, let us take the Matsubara sums.  The standard technique leads  to the following useful formula,
\begin{align}
  &T\sum_{m=-\infty}^\infty \frac{g(\ri\nu_m+\mu)}{[(\ri\nu_m+\mu)^2-E_1^2][(\ri\nu_m+\mu-\ri\omega_n)^2-E_2^2]} \notag\\
  &= -\sum_{s_1,s_2=\pm1}\frac{s_1s_2}{4E_1E_2}\frac{1}{-\ri\omega_n+s_1E_1-s_2E_2} \times \notag\\
  &\qquad\qquad\times\Big[
  n_\mathrm{F}(\mu-s_1E_1)g(s_1E_1) - n_\mathrm{F}(\mu-s_2E_2)g(s_2E_2+\ri\omega_n) \Bigr]\,.
\end{align}
Here $g(\ri\nu_m+\mu)$ is an arbitrary function satisfying $\lim_{z\to\infty}|g(z)|<|z|^3$.  Using this, we can simplify the polarization tensor in the following form:
\begin{equation}
\begin{split}
 &\Pi^{\mu\nu}(\ri\omega_n,\bm{k})\\
 &=
  \sum_{k,l=0}^\infty  \sum_{s_1,s_2=\pm1}e^2 \int \frac{\dd{p_z}}{2\pi}\frac{s_1s_2}{4E_{k}E_{l}}\frac{1}{-\ri\omega_n+s_1E_{k}-s_2E_{l}} \times\\
&\quad\times \Bigl[
n_\mathrm{F}(\mu-s_1E_{k})\LepT_{kl}^{\mu\nu}(s_1E_{k},p_z;k_\parallel,z)  -n_\mathrm{F}(\mu-s_2E_{l})\LepT_{kl}^{\mu\nu}(s_2E_{l}+\ri\omega_n,p_z;k_\parallel,z)
\Bigr]\,.
\end{split}
\end{equation}
Unless confusion arises, we omit the momentum arguments for the energy for notational brevity, that is,
\begin{equation}
  E_k=E_k(p_z), \qquad E_l=E_l(p_z-k_z)\,,
\end{equation}
in our convention.

It is a well-established procedure to obtain the retarded function by performing analytical continuation, $\ri\omega_n\to k_0+\ri\epsilon$, i.e.,
\begin{equation}
\begin{split}
  & \Pi_R^{\mu\nu}(k) =
  \sum_{k,l=0}^\infty \sum_{s_1,s_2=\pm1} e^2\int \frac{\dd{p_z}}{2\pi}\frac{s_1s_2}{4E_{k}E_{l}}\frac{1}{-k_0+s_1E_{k}-s_2E_{l}-\ri\epsilon} \times\\
  &\qquad\times \Bigl[
n_\mathrm{F}(\mu-s_1E_{k})\LepT_{kl}^{\mu\nu}(s_1E_{k},p_z;k_\parallel,z)-n_\mathrm{F}(\mu-s_2E_{l})\LepT_{kl}^{\mu\nu}(s_2E_{l}+k_0,p_z;k_\parallel,z) 
\Bigr]\,.
\end{split}
\end{equation}
Thanks to Dirac's delta function, it is far easier to evaluate the imaginary part,
\begin{equation}
  \begin{split}\label{eq:imPiR}
  2\Im\Pi_R^{\mu\nu}(k) &=
  \sum_{k,l=0}^\infty  \sum_{s_1,s_2=\pm1}e^2 \int \frac{\dd{p_z}}{2\pi}\frac{s_1s_2}{4E_{k}E_{l}}(2\pi)\delta(k_0-s_1E_{k}+s_2E_{l}) \times\\
  &\quad\times \Bigl[
n_\mathrm{F}(\mu-s_1E_{k})-n_\mathrm{F}(\mu-s_2E_{l})\Bigr] \LepT_{kl}^{\mu\nu}(s_1E_{k},p_z;k_\parallel,z)\,.
\end{split}
\end{equation}
Here, for the second term in the square brackets, we replaced $s_2 E_l+k_0$ by $s_1 E_k$ in $\LepT_{kl}^{\mu\nu}$ using the Dirac's delta function constraint. 
We note that the imaginary part at finite temperature and density has already been calculated in ref.~\cite{Wang:2021eud}.
In this literature, the fermion propagator takes the mixed coordinate-momentum representation, and the integration of the transverse direction is handled ingeniously in coordinate space.
Alternatively, we now conduct the straightforward calculation in a common prescription in terms of the momentum-space propagator.

Once the imaginary part is given, the real part is reconstructed by the Kramers-Kronig relation:
\begin{equation}\label{eq:Kramers-Kronig}
  \Re\Pi_R^{\mu\nu}(k) =
  \mathop{\mathcal{P}} \int \frac{\dd{s}}{2\pi} \frac{\mathop{2\Im}\Pi_R^{\mu\nu}(s,\bm{k})}{s - k_0}\,,
\end{equation}
where $\mathop{\mathcal{P}}$ denotes the principal value.
When this integral diverges, a renormalization procedure will be required.

%%%%%%%%%
\subsection{Solving the energy conservation constraint}
Before going into the evaluation of $\LepT^{\mu\nu}_{kl}$, we shall review the solution of the energy conservation law, which was already discussed in refs.~\cite{Baier:2007dw, Hattori:2012je, Hattori:2012ny, Wang:2020dsr, Wang:2021ebh, Wang:2021eud}.
The Dirac's delta function $\delta(k_0-s_1E_{k}+s_2E_{l})$ in eq.~\eqref{eq:imPiR} gives the energy conservation relation $k_0-s_1E_{k}+s_2E_{l}=0$ and limits the possible values of $p_z$.

From $k_0-s_1E_{k}+s_2E_{l}=0$, the $z$-components of the fermions' momentum must have the following specific values:
\begin{equation}
\begin{split}
\label{eq:p_zt}
p_{z\pm} =\frac{k_z}{2}a_{kl}
\pm\frac{k_0}{2}b_{kl}\,,
\end{split}
\end{equation}
where
\begin{align}
a_{kl}&\coloneqq1+\frac{m_k^2-m_l^2}{k_\parallel^2}\,, \label{eq:akl}\\
b^2_{kl}&\coloneqq a_{kl}^2-\frac{4m_k^2}{k_\parallel^2}
=\left[1-\frac{(m_{k}+m_{l})^2}{k_\parallel^2}\right]\left[1-\frac{(m_{k}-m_{l})^2}{k_\parallel^2} \right]\,. 
\end{align}
The corresponding energies are 
\begin{align}
E_{k\pm}&=\sqrt{p_{z\pm}^2+m_k^2}
=\frac{1}{2}\left|{k_0}a_{kl}
\pm { k_z}b_{kl}\right|\,,\\
E_{l\mp}&=\sqrt{(p_{z\pm}-k_z)^2+m_l^2}=\frac{1}{2}\left|{k_0}a_{lk}
\mp{ k_z}b_{kl}\right|\,.
\end{align}
We note that the solution does not exist for some values of $k_\parallel^2$.
At least, $k_\parallel^2\geq(m_{k}+m_l)^2$ or $k_\parallel^2\leq(m_{k}-m_l)^2$ are necessary for a real $b_{kl}$ to exist.

For the $p_z$-integration in eq.~\eqref{eq:imPiR}, we need the following derivative:
\begin{equation}
\begin{split}
\left.\frac{\dd}{\dd{p_z}}(k_0-s_1E_{k}+s_2E_{l})\right|_{p_z=p_{z\pm}}
=\left.\frac{-p_z}{s_1E_k} +\frac{p_z-k_z}{s_2E_l}\right|_{p_z=p_{z\pm}}
=\pm\frac{ k_\parallel^2}{2s_1s_2E_{k\pm}E_{l\mp}}b_{kl}\,,
\end{split}
\end{equation}
where we used $k_0p_{z\pm} -s_1k_zE_{k\pm}=\pm k_\parallel^2b_{kl}/2$ in the end, which can be derived from the energy conservation relation, $k_0-s_1E_{k}+s_2E_{l}=0$.

For $k_0-E_{k}-E_{l}=0$ ($s_1=1$, $s_2=-1$), this process corresponds to the decay process.
The solution exists if $k_{\parallel}^2\geq(m_k+m_l)^2$ and $k_0>0$.
Therefore, the Dirac's delta function becomes
\begin{equation}
\delta(k_0-E_{k}-E_{l})=\sum_{t=\pm}\frac{2E_{k,t}E_{l,-t}}{b_{kl}k_\parallel^2}\delta(p_z-p_{zt}) 
\theta\bigl(k_\parallel^2-(m_k+m_l)^2\bigr)\theta(k_0)\,.
\end{equation}
For given $k_\parallel^2$, there is the upper limit  of $k$ and $l$,
which is obtained by the maximum $n$ satisfying $k_\parallel^2\geq(m_0+m_{n})^2$.
The solution is $n_\text{max} = \lfloor(k_\parallel^2 - \sqrt{4m^2k_\parallel^2})/(2eB)\rfloor$,
where $\lfloor x \rfloor=\max\{n \in\mathbb{Z}| n\leq x\}$ is the floor function.

For $k_0+E_{k}+E_{l}=0$ ($s_1=-1$, $s_2=1$), this is a similar case to $k_0-E_{k}-E_{l}=0$. 
We obtain
\begin{equation}
\delta(k_0+E_{k}+E_{l})=\sum_{t=\pm}\frac{2E_{k,t}E_{l,-t}}{b_{kl}k_\parallel^2}\delta(p_z-p_{zt})
\theta\bigl(k_\parallel^2-(m_k+m_l)^2\bigr)\theta(-k_0)\,.
\end{equation}

The situation for $k_0-E_{k}+E_{l}=0$ ($s_1=s_2=1$) is a bit different from the previous two cases. 
We will consider the timelike region and the spacelike region separately. 
Let us begin with the timelike region, i.e., $0<k_\parallel^2\leq(m_k-m_l)^2$.
The solution exists when $m_l<m_k$ for $k_0>0$ and $m_l>m_k$ for $k_0<0$.
In the former case, there is the minimum $k$, which is $k_\text{min} = \lceil (k_\parallel^2 + \sqrt{4m^2k_\parallel^2})/(2eB)\rceil$,
where $\lceil x \rceil=\min\{n \in\mathbb{Z}| x\leq n\}$ is the ceiling function.
The Dirac's delta function reduces to 
\begin{equation}
\theta(k_\parallel^2)\delta(k_0-E_{k}+E_{l})=\sum_{t=\pm}\frac{2E_{k,t}E_{l,-t}}{b_{kl}k_\parallel^2}\delta(p_z-p_{zt})
\theta(k_\parallel^2)\theta\bigl((m_k-m_l)^2-k_\parallel^2\bigr)\theta\bigl(k_0(m_k-m_l)\bigr)\,.
\end{equation}
Next, consider the spacelike region, $k_\parallel^2<0$.
In this case, $p_z=p_{z+}$ for $k_z>0$ or $p_z=p_{z-}$ for $k_z<0$ is the solution. 
Therefore, the Dirac's delta function for $k_\parallel^2<0$ becomes
\begin{equation}
\begin{split}
\theta(-k_\parallel^2)\delta(k_0-E_{k}+E_{l}) = -\sum_{t=\pm}\frac{2E_{k,t}E_{l,-t}}{b_{kl}k_\parallel^2}\delta(p_z-p_{zt})\theta(tk_z)
\theta(-k_\parallel^2)\,.
\end{split}
\end{equation}
In summary, the Dirac's delta function is
\begin{equation}
\begin{split}
\delta(k_0-E_{k}+E_{l}) &= \sum_{t=\pm}\frac{2E_{k,t}E_{l,-t}}{b_{kl}k_\parallel^2}\delta(p_z-p_{zt})\times\\
&\quad\times \theta\bigl((m_k-m_l)^2-k_\parallel^2\bigr)\Bigl[\theta(k_\parallel^2)\theta\bigl(k_0(m_k-m_l)\bigr)
-\theta(tk_z)\theta(-k_\parallel^2)\Bigr]\,.
\end{split}
\end{equation}

The remaining situation, $k_0+E_{k}-E_{l}=0$ ($s_1=s_2=-1$), is similar to the previous case. We may change $k_0$ to $-k_0$. Then, the Dirac's delta function is
\begin{equation}
\begin{split}
\delta(k_0+E_{k}-E_{l}) &= \sum_{t=\pm}\frac{2E_{k,t}E_{l,-t}}{b_{kl}k_\parallel^2}\delta(p_z-p_{zt})\times\\
&\quad\times \theta\bigl((m_k-m_l)^2-k_\parallel^2\bigr)\Bigl[\theta(k_\parallel^2)\theta\bigl(k_0(m_l-m_k)\bigr)
-\theta(tk_z)\theta(-k_\parallel^2)\Bigr]\,.
\end{split}
\end{equation}

\subsection{Polarization tensor with convenient basis}

The tensor indices, $\mu$, $\nu$, refer to the directions in the coordinate system, and it would be more convenient to consider the tensor structure relative to the photon momentum $k^\mu$ and the field strength tensor $F^{\mu\nu}$ with $\bB$ along the $z$ axis.  This means that $F^{\mu\nu}=-\epsilon_\perp^{\mu\nu}B$ with $\epsilon_\perp^{12}=-\epsilon_\perp^{21}=+1$ and others are zero.
The dual tensor is $\tilde{F}^{\mu\nu}=\frac{1}{2}\epsilon^{\mu\nu\rho\sigma}F_{\rho\sigma}=\epsilon_\parallel^{\mu\nu}B$ with $\epsilon_\parallel^{03}=-\epsilon_\parallel^{30}=+1$ and others are zero.  We note $\epsilon^{0123}=-1$ in our convention.
For the new tensor basis, $k^\mu$ itself is one natural choice of the reference vector, and three vectors orthogonal to $k^\mu$ are chosen as
\begin{equation}
  \beta_1^\mu \coloneqq \frac{k^2 k_\perp^\mu-k_\perp^2 k^\mu}{\sqrt{-k_\perp^2 k^2 k_\parallel^2}}\,, \qquad
  \beta_2^\mu \coloneqq \frac{F^{\mu\nu}k_\nu}{B \sqrt{-k_\perp^2}} = \frac{-\epsilon_{\perp}^{\mu\nu}k_\nu}{\sqrt{-k_\perp^2}}\,, \qquad
  \beta_3^\mu \coloneqq \frac{\tilde{F}^{\mu\nu}k_\nu}{B\sqrt{k_\parallel^2}} =
  \frac{\epsilon_{\parallel}^{\mu\nu}k_\nu}{\sqrt{k_\parallel^2}}\,.
\label{eq:tensordecomposition}
\end{equation}
This tensor basis was already discussed in some literature~\cite{Baier:2007dw, Hattori:2012je, Hattori:2012ny, Wang:2020dsr, Wang:2021ebh, Wang:2021eud}, and we use the same notation $\beta_i^\mu$ as defined in ref.~\cite{Baier:2007dw}.
We can easily check that $\beta^\mu_i$'s satisfy:
\begin{equation}
  k\cdot \beta_i=0,\,
  \qquad
  {\beta}_i \cdot{\beta}_j = -\delta_{ij}
  \label{eq:beta_orth}.
\end{equation}
The linear polarization along $\beta_2^\mu$ corresponds to the X-mode, i.e., the mode perpendicular to the plane spanned by $\bk$ and $\bB$ ($\bk$-$\bB$ plane).  Then, there are two independent vectors to span the $\bk$-$\bB$ plane.  Among them, $\beta_3^\mu$ is directed to $\bB$ corresponding to the O-mode, while $\beta_1^\mu$ is uniquely defined from $\beta_2^\mu$, $\beta_3^\mu$, $k^\mu$ using eq.~\eqref{eq:beta_orth}.

Moreover, they form a projection operator,
\begin{equation}
  \sum_{i=1}^3 \beta_{i}^{\mu} \beta_{i}^{\nu} = \frac{k^{\mu}k^{\nu}}{k^2}-\eta^{\mu\nu}\,.
\end{equation}

Armed with these basis vectors, we can express the inverse of Green's function of the photon, which is defined by the polarization tensor as
\begin{equation}
  [G_R^{-1}(k)]^{\mu\nu} \coloneqq \eta^{\mu\nu}k^2-k^\mu k^\nu -\Pi_R^{\mu\nu}(k)\,.
\end{equation}
The Ward-Takahashi identity implies $[G_R^{-1}(k)]^{\mu\nu} k_\mu = [G_R^{-1}(k)]^{\mu\nu} k_\nu = 0$.  These conditions require $\Pi_R^{\mu\nu}(k) k_\mu= \Pi_R^{\mu\nu}(k) k_\nu=0$, leading to a decomposition of
\begin{equation}
  \Pi_R^{\mu\nu}(k) \eqqcolon \beta_i^\mu\beta_j^\nu \tilde{\Pi}_R^{ij}\,,
\end{equation}
or equivalently $\tilde{\Pi}_R^{ij} = \Pi_R^{\mu\nu}\beta_{i\mu} \beta_{j\nu}$.  In this way, we define the tensors in this basis as
\begin{equation}
    \tilde{\LepTdensity}_{kl}^{ij} \coloneqq \LepTdensity_{kl}^{\mu\nu} \beta_{i\mu}\beta_{j\nu}\,,\qquad
    \tilLepT_{kl}^{ij} \coloneqq \LepT_{kl}^{\mu\nu} \beta_{i\mu}\beta_{j\nu}\,.
\end{equation}

With this basis, considering the $p_z$ integration under the energy conservation, we finally reach
\begin{align}
\label{eq:2imPi}
&2\Im\tilde{\Pi}_R^{ij}(k) \notag\\
 &= \sum_{k,l=0}^\infty  \sum_{t=\pm}\frac{e^2\tilLepT_{kl}^{ij}(p_\parallel, k)\bigr|_{p_z=p_{zt}}}{2k_\parallel^2b_{kl}}\Bigl\{\notag\\
& \quad-\bigl[
n_\mathrm{F}(E_{k,t}-\mu)-n_\mathrm{F}(E_{l,-t}-\mu)\bigr]
\theta\bigl((m_k-m_l)^2-k_\parallel^2\bigr)\bigl[\theta(k_\parallel^2)\theta\bigl(k_0(m_k-m_l)\bigr)
-\theta(tk_z)\theta(-k_\parallel^2)\bigr]+\notag\\
&\quad+\bigl[
n_\mathrm{F}(E_{k,t}+\mu)-n_\mathrm{F}(E_{l,-t}+\mu)\bigr]
\theta\bigl((m_k-m_l)^2-k_\parallel^2\bigr)\bigl[\theta(k_\parallel^2)\theta\bigl(k_0(m_l-m_k)\bigr)
-\theta(tk_z)\theta(-k_\parallel^2)\bigr]-\notag\\
&\quad - \bigl[
1-n_\mathrm{F}(E_{k,t}-\mu)-n_\mathrm{F}(E_{l,-t}+\mu)\bigr]\theta(k_0)\theta\bigl(k_\parallel^2-(m_l+m_k)^2\bigr)+\notag\\
&\quad + \bigl[
1-n_\mathrm{F}(E_{k,t}+\mu)-n_\mathrm{F}(E_{l,-t}-\mu)\bigr]\theta(-k_0)\theta\bigl(k_\parallel^2-(m_l+m_k)^2\bigr)
\Bigr\}\,.
\end{align}
It should be noted that we simplified the notation from $\tilde \LepT_{kl}^{ij}(p_0, p_z=p_{zt};k_\parallel, z)$ to $\tilde \LepT_{kl}^{ij}(p_\parallel;k)|_{p_z=p_{zt}}$.
We also use the Kramers-Kronig relation eq.~\eqref{eq:Kramers-Kronig} for the calculation of the real part of the polarization tensor.

%%%%%

%%%%%%%%%%
\section{Analytical evaluation of the lepton tensor integral}

For the rest of this paper, we will complete the full analytical evaluation of $\LepT_{kl}^{\mu\nu}$.
Actually, this quantity was computed in ref.~\cite{Wang:2021eud} by completing the integration ingeniously with respect to the transverse coordinates.
Here, we will derive the equivalent results using the more straightforward method according to the common knowledge of quantum field theory, i.e., we adopt the momentum representation and directly perform the transverse integration in eq.~\eqref{eq:h-kl}.  Once the analytical form of $\LepT_{kl}^{\mu\nu}$ is known, we can numerically take the sum over the Landau levels with respect to $k,l$.

%%%%%
\subsection{Taking the Dirac trace}

Let us simplify the explicit forms of $\tilde{\LepTdensity}_{kl}^{ij}$ here.  In this procedure, we do not have to perform any integration, and the analytical manipulations are straightforward.

First, we shall write down $\LepTdensity_{kl}^{\mu\nu}(p,k)$ as follows:
\begin{equation}
  \begin{split}
  & \LepTdensity_{kl}^{\mu\nu}(p,k) = \tr\Bigl\{
  \gamma^\mu
  \Bigl[ (\slashed{p}_\parallel+m)(P_+ A_{k+}(\xi_p) + P_- A_{k-}(\xi_p)) + \slashed{p}_\perp B_k(\xi_p) \Bigr] \gamma^\nu \times\\
  & \quad\times \Bigl[ (\slashed{p}_\parallel-\slashed{k}_\parallel+m)(P_+ A_{l+}(\xi_{p-k}) +P_- A_{l-}(\xi_{p-k}))+(\slashed{p}_\perp-\slashed{k}_\perp) B_l(\xi_{p-k})\Bigr]
  \Bigr\} \,.
  \end{split}
\end{equation}
There appear to be nine different cross-terms.  We can complete elementary but tedious calculations of the trace, such as
\begin{equation}
  \begin{split}
  &\tr\bigl[\gamma^{\mu}(\slashed{p}_\parallel + m)P_{\pm} \gamma^\nu
  (\slashed{p}_\parallel - \slashed{k}_\parallel + m)P_{\pm}\bigr] \\
  &\qquad\qquad = 2\bigl[
(p_\parallel-k_\parallel)^\mu p_\parallel^\nu+(p_\parallel-k_\parallel)^\nu p_\parallel^\mu
+(m^2-(p_\parallel-k_\parallel)\cdot p_\parallel )\eta_\parallel^{\mu\nu}\bigr]\,.
  \end{split}
\end{equation}
Here, the longitudinal metric is introduced $p_\parallel^\mu = \eta_\parallel^{\mu\nu}p_\nu$ for any $p^\mu$ and the transverse metric is $\eta_\perp^{\mu\nu}=\eta^{\mu\nu}-\eta_\parallel^{\mu\nu}$ accordingly.  The transverse one is found in a similar trace,
\begin{equation}
  \tr\bigl[\gamma^{\mu} (\slashed{p}_\parallel + m) P_{\mp} \gamma^\nu (\slashed{p}_\parallel - \slashed{k}_\parallel + m)P_{\pm} \bigr]
  = 2(m^2 -(p_\parallel-k_\parallel)\cdot p_\parallel)(\eta_\perp^{\mu\nu}\mp \ri \epsilon^{\mu\nu}_\perp)\,.
\end{equation}
Another nontrivial trace is
\begin{equation}
  \tr\bigl[ \gamma^{\mu} \slashed{p}_\perp \gamma^\nu (\slashed{p}_\parallel-\slashed{k}_\parallel+m)P_\pm \bigr]
  = 2(p_\parallel - k_\parallel)^\mu (p_\perp^\nu\pm\ri\epsilon_\perp^{\nu\rho}p_{\perp\rho})
  +2(p_\parallel - k_\parallel)^\nu (p_\perp^\mu\mp\ri\epsilon_\perp^{\mu\rho}p_{\perp\rho})\,.
\end{equation}
Other cross terms can be recovered by exchanging the momenta and the indices.  After all, the explicit expression is 
\begin{equation}
\label{eq:Hkl}
\begin{split}
&\LepTdensity_{kl}^{\mu\nu}(p,k)\\
&=
2\Bigl[
(p_\parallel-k_\parallel)^\mu p_\parallel^\nu+(p_\parallel-k_\parallel)^\nu p_\parallel^\mu
+(m^2-(p_\parallel-k_\parallel)\cdot p_\parallel )\eta_\parallel^{\mu\nu}\Bigr]\times\\
&\qquad \times\Bigl[A_{k+}(\xi_p)A_{l+}(\xi_{p-k})
+A_{k-}(\xi_p)A_{l-}(\xi_{p-k})\Bigr]+\\
&\quad+
2\Bigl[m^2-(p_\parallel-k_\parallel)\cdot p_\parallel\Bigr]
\eta_\perp^{\mu\nu}\Bigl[A_{k-}(\xi_p)A_{l+}(\xi_{p-k})
+A_{k+}(\xi_p)A_{l-}(\xi_{p-k})
\Bigr]-\\
&\quad
-2\Bigl[m^2-(p_\parallel-k_\parallel)\cdot p_\parallel\Bigr]\ri \epsilon^{\mu\nu}_\perp\Bigl[A_{k-}(\xi_p)A_{l+}(\xi_{p-k})
-A_{k+}(\xi_p)A_{l-}(\xi_{p-k})
\Bigr]+
\\
&\quad+4\Bigl[(p_\perp-k_\perp)^\mu p_\perp^\nu+(p_\perp-k_\perp)^\nu p_\perp^\mu - \eta^{\mu\nu}(p_\perp-k_\perp)\cdot p_\perp\Bigr]
 B_k(\xi_p)B_l(\xi_{p-k})+\\
&\quad+2\Bigl[
(p_\parallel-k_\parallel)^\mu p_\perp^\nu 
+ (p_\parallel-k_\parallel)^\nu p_\perp^\mu
\Bigr]\Bigl[A_{l+}(\xi_{p-k})
+A_{l-}(\xi_{p-k})
\Bigr]B_k(\xi_p)+\\
&\quad+2\Bigl[
 (p_\parallel-k_\parallel)^\mu\ri\epsilon_\perp^{\nu\rho}p_{\perp\rho}
- (p_\parallel-k_\parallel)^\nu\ri\epsilon_\perp^{\mu\rho}p_{\perp\rho}
\Bigr]\Bigl[A_{l+}(\xi_{p-k})
-A_{l-}(\xi_{p-k})\Bigr]B_k(\xi_p)+
\\
&\quad+2\Bigl[
(p_\perp-k_\perp)^\mu p_\parallel^\nu
+(p_\perp-k_\perp)^\nu p_\parallel^\mu
\Bigr]\Bigl[A_{k+}(\xi_p)
+A_{k-}(\xi_p)
\Bigr]B_l(\xi_{p-k})+\\
&\quad+2\Bigl[
(p_{\perp}-k_{\perp})_\rho\ri\epsilon_\perp^{\mu\rho}p_\parallel^\nu
-(p_\perp-k_\perp)_\rho\ri\epsilon_\perp^{\nu\rho}p_\parallel^\mu
\Bigr]\Bigl[A_{k+}(\xi_p)
-A_{k-}(\xi_p)
\Bigr]B_l(\xi_{p-k})\,.
\end{split}
\end{equation}

Next, let us transform $\LepTdensity_{kl}^{\mu\nu}$ into $\tilde{\LepTdensity}_{kl}^{ij} = \beta_{i\mu}\beta_{j\nu }\LepTdensity_{kl}^{\mu\nu}$.  In view of the original definition~\eqref{eq:tensordecomposition}, $\beta_1^\mu$ contains a term $\sim k^\mu$, but $k_\mu \Pi_R^{\mu\nu}=0$ thanks to the Ward-Takahashi identity.  Thus, we can safely drop this term and replace $\beta_{1}^\mu$ by $k^2 k_\perp^\mu/\sqrt{-k_\perp^2 k^2 k_\parallel^2}$ to calculate $\tilde{\LepTdensity}_{kl}^{ij}$.

We take a look at the diagonal components and then proceed to the off-diagonal components.  For $i=j=1$, after some calculations, we find
\begin{align}
  & \tilde{\LepTdensity}_{kl}^{11}(p,k) = \frac{k^2}{k_\parallel^2}\biggl\{ \bigl[ p_\parallel^2 + (p_\parallel-k_\parallel)^2 - 2m^2 - k_\parallel^2 \bigr]
  \bigl[A_{k-}(\xi_p) A_{l+}(\xi_{p-k}) + A_{k+}(\xi_p)A_{l-}(\xi_{p-k}) \bigr] + \notag\\
  &\qquad\qquad + \frac{4}{\bk_\perp^2} \bigl[ 2(\bp_\perp-\bk_\perp)\cdot \bk_\perp ( \bp_\perp\cdot \bk_\perp) - \bk_\perp^2(\bp_\perp-\bk_\perp)\cdot \bp_\perp \bigr] B_k(\xi_p)B_l(\xi_{p-k}) \biggl\}\,.
\end{align}
When we evaluate $2\Im\Pi_R$ in eq.~\eqref{eq:imPiR}, we can use the on-shell conditions, i.e., $p_\parallel^2-2eBk=m^2$ and $(p_\parallel-k_\parallel)^2-2eBl=m^2$.  These conditions slightly simplify the above expression to
\begin{equation}
\label{eq:h11_onshell}
  \begin{split}
  &\tilde{\LepTdensity}_{kl}^{11}(p,k) = \frac{k^2}{k_\parallel^2}\biggl\{
 \bigl[ 2eB(k+l)-k_\parallel^2 \bigr] \bigl[ A_{k-}(\xi_p) A_{l+}(\xi_{p-k})+ A_{k+}(\xi_p) A_{l-}(\xi_{p-k})\bigr] -\\
  &\qquad\qquad\qquad -4\biggl[ \bp_\perp^2-2\frac{(\bp_\perp\cdot \bk_\perp)^2}{\bk_\perp^2} + ( \bp_\perp\cdot \bk_\perp) \biggr]
  B_k(\xi_p)B_l(\xi_{p-k}) \biggr\}\,.
  \end{split}
\end{equation}
Similarly, using the same on-shell conditions for the other diagonal components, we get
\begin{equation}
\label{eq:h22_onshell}
\begin{split}
\tilde{\LepTdensity}_{kl}^{22}(p,k)&=
\bigl[2eB(k+l)-k_\parallel^2\bigr]
\bigl[A_{k-}(\xi_p)A_{l+}(\xi_{p-k})
+A_{k+}(\xi_p)A_{l-}(\xi_{p-k})
\bigr]+\\
&\quad+{4}\biggl[\bm{p}_\perp^2 - 2\frac{(\bm{p}_\perp\cdot \bm{k}_\perp)^2}{\bm{k}_\perp^2} +(\bm{p}_\perp\cdot \bm{k}_\perp)\biggr]
 B_k(\xi_p)B_l(\xi_{p-k})\,,
\end{split}
\end{equation}
and
\begin{align}
\label{eq:h33_onshell}
&\tilde{\LepTdensity}_{kl}^{33}(p,k) \notag\\
&=\frac{1}{k_\parallel^2}\Bigl\{
 \bigl[2eB(k-l)\bigr]^2
-k_\parallel^2\bigl[4m^2+2eB(k+l)\bigr]\Bigr\}\bigl[A_{k+}(\xi_p)A_{l+}(\xi_{p-k})
+A_{k-}(\xi_p)A_{l-}(\xi_{p-k})\bigr]- \notag\\
&\quad -4(\bm{p}_\perp-\bm{k}_\perp)\cdot \bm{p}_\perp
 B_k(\xi_p)B_l(\xi_{p-k})\,.
\end{align}

Next, we move on to the off-diagonal components.
Since $\tilde \LepTdensity^{ij}_{kl}(p,k)$ is Hermitian, there are only three off-diagonal components that have to be calculated.
First computing the component $\tilde \LepTdensity^{12}_{kl}(p,k)$, the expression is simplified to
\begin{align}
\label{eq:H12}
&\tilde{\LepTdensity}^{12}_{kl}(p,k) \notag\\
&=\frac{k^2}{\sqrt{k^2k_\parallel^2}}\biggl\{
-\ri\bigl[p_\parallel^2+(p_\parallel- k_\parallel)^2-2m^2-k_\parallel^2\bigr]\bigl[A_{k-}(\xi_p)A_{l+}(\xi_{p-k})
-A_{k+}(\xi_p)A_{l-}(\xi_{p-k})
\bigr]+ \notag\\
&\quad+4\frac{\epsilon_{\perp ij}p_\perp^i k_\perp^j}{\bm{k}_\perp^2}
\bm{k}_{\perp}\cdot(2\bm{p}_\perp-\bm{k}_\perp)
 B_k(\xi_p)B_l(\xi_{p-k})
\biggr\}\,.
\end{align}
After the integration of $p_\perp^i \bm{k}_{\perp}\cdot(2\bm{p}_\perp-\bm{k}_\perp) B_k(\xi_p)B_l(\xi_{p-k})$ with respect to $p_\perp$ for the evaluation of $\tilde \LepT^{12}_{kl}$, the result is proportional to $k_\perp^i$, so that the final result vanishes because of $\epsilon_{\perp ij} k^i k^j=0$.
Therefore, we can drop the last term in eq.~\eqref{eq:H12}.
Moreover, we can use the on-shell conditions, which are used in the computation of the diagonal components.
Finally, $\tilde{\LepTdensity}^{12}_{kl}(p,k)$ reduces to
\begin{align}
\tilde{\LepTdensity}^{12}_{kl}(p,k)=-\ri \frac{k^2}{\sqrt{k^2k_\parallel^2}}
\bigl[2eB(k+l)-k_\parallel^2\bigr]\bigl[A_{k-}(\xi_p)A_{l+}(\xi_{p-k})
-A_{k+}(\xi_p)A_{l-}(\xi_{p-k})
\bigr]\,.
\end{align}
Next, the other off-diagonal components are
\begin{equation}
\begin{split}
\label{eq:H13}
\tilde{\LepTdensity}^{13}_{kl}(p,k)&=\frac{- p_\parallel^\nu\epsilon_{\parallel\nu\lambda}k^\lambda}{k_\parallel^2}\frac{k^2}{\sqrt{-k_\perp^2k^2}}\Bigl\{2(
\bm{p}_{\perp}\cdot  \bm{k}_\perp)\bigl[A_{l+}(\xi_{p-k})
+A_{l-}(\xi_{p-k})
\bigr]B_k(\xi_p)+\\
&\quad+2\bigl[
(\bm{p}_\perp-\bm{k}_\perp)\cdot\bm{k}_{\perp}
\bigr]\bigl[A_{k+}(\xi_p)
+A_{k-}(\xi_p)\bigr]B_l(\xi_{p-k})\Bigr\}\,,
\end{split}
\end{equation}
and
\begin{equation}
\begin{split}
\label{eq:H23}
\tilde{\LepTdensity}^{23}_{kl}(p,k)&=\frac{-\ri p_\parallel^\nu\epsilon_{\parallel \nu\lambda}k^\lambda}{\sqrt{-k_\perp^2k_\parallel^2}}\Bigl\{2(\bm{p}_{\perp}\cdot \bm{k}_\perp)\bigl[A_{l+}(\xi_{p-k})
-A_{l-}(\xi_{p-k})\bigr]B_k(\xi_p)-
\\&\quad-2\bigl[(\bm{p}_{\perp}-\bm{k}_{\perp})\cdot\bm{k}_\perp
\bigr]\bigl[A_{k+}(\xi_p)
-A_{k-}(\xi_p)
\bigr]B_l(\xi_{p-k})
\Bigr\}\,.
\end{split}
\end{equation}
Equations~\eqref{eq:H13} and \eqref{eq:H23} contain a common factor, $p_\parallel^\nu\epsilon_{\parallel \nu\lambda}k^\lambda$, which obviously depends on $p_z$.
For the imaginary part, $p_z$ takes two values of $p_{zt=\pm}$ defined in eq.~\eqref{eq:p_zt}.
Therefore, $\tilde \LepTdensity^{13}_{kl}(p,k)$ and $\tilde \LepTdensity^{23}_{kl}(p,k)$ depend on $t$.
To make it explicit, we introduce the following notation:
\begin{align}
    \tilde \LepTdensity^{13}_{kl}(p,k;t)\coloneqq \tilde \LepTdensity^{13}_{kl}(p,k)\bigr|_{p_z=p_{zt}}\,,\qquad\tilde \LepTdensity^{23}_{kl}(p,k;t)\coloneqq \tilde \LepTdensity^{23}_{kl}(p,k)\bigr|_{p_z=p_{zt}}\,.
\end{align}
The energy conservation law $p_\parallel^\nu\epsilon_{\parallel \nu\lambda}k^\lambda=tk_\parallel^2b_{kl}/2$ simplifies eqs.~\eqref{eq:H13} and \eqref{eq:H23} to
\begin{equation}
\begin{split}
\tilde{\LepTdensity}^{13}_{kl}(p,k;t)&=-tb_{kl}\frac{k^2}{\sqrt{\bm{k}_\perp^2k^2}}\Bigl\{(\bm{p}_{\perp}\cdot  \bm{k}_\perp)\bigl[A_{l+}(\xi_{p-k})
+A_{l-}(\xi_{p-k})
\bigr]B_k(\xi_p)+\\
&\quad+\bigl[(\bm{p}_\perp-\bm{k}_\perp)\cdot\bm{k}_{\perp}
\bigr]\bigl[A_{k+}(\xi_p)
+A_{k-}(\xi_p)
\bigr]B_l(\xi_{p-k})
\Bigr\}\,,
\end{split}
\end{equation}
and
\begin{equation}
\begin{split}
\tilde{\LepTdensity}^{23}_{kl}(p,k;t)&=-\ri tb_{kl}\frac{k_\parallel^2}{ \sqrt{\bm{k}_\perp^2k_\parallel^2}}\Bigl\{(\bm{p}_{\perp}\cdot \bm{k}_\perp)\bigl[A_{l+}(\xi_{p-k})
-A_{l-}(\xi_{p-k})\bigr]B_k(\xi_p)-
\\
&\quad-\bigl[(\bm{p}_{\perp}-\bm{k}_{\perp})\cdot\bm{k}_\perp
\bigr]\bigl[A_{k+}(\xi_p)
-A_{k-}(\xi_p)
\bigr]B_l(\xi_{p-k})
\Bigr\}\,.
\end{split}
\end{equation}
Note that we omit to write down the explicit expression of $\tilde \LepTdensity^{ij}_{kl}(p,k)$ ($i>j$) because of the Hermitian property of $\tilde \LepTdensity^{ij}_{kl}(p,k)$.

%%%%%
\subsection{Transverse momentum integration}
In order to present the explicit form of $\tilde \LepT^{ij}_{kl}(p_\parallel, k)$, it is necessary to conduct the following integration:
\begin{equation}
    \tilde \LepT^{ij}_{kl}(p_\parallel, k)=\int\dfrac{\dd[2]{p_\perp}}{(2\pi)^2}\tilde \LepTdensity^{ij}_{kl}(p,k)\,.
\end{equation}
Since this integral contains some complicated integrations, we give several formulas first:
\begin{equation}
\label{eq:Akl}
    \begin{split}
        \Ffunc_{k,l}(z)&\coloneqq\int \frac{\dd[2]{p_\perp}}{(2\pi)^2} \re^{-2\xi_p}L_{k}(4\xi_p)
        \re^{-2\xi_{p-k}}L_{l}(4\xi_{p-k})\\
        &=\frac{eB}{8\pi}\re^{-z}\frac{k!}{l!}(-z)^{l-k}\Bigl[L_{k}^{(l-k)}(z)\Bigr]^2\,,
    \end{split}
\end{equation}
\begin{equation}
\label{eq:Bkl}
    \begin{split}
        \Gfunc_{k,l}(z)& \coloneqq\int
        \frac{\dd[2]{p_\perp}}{(2\pi)^2} \re^{-2\xi_p}L^{(1)}_{k-1}(4\xi_p)
        \re^{-2\xi_{p-k}}L^{(1)}_{l-1}(4\xi_{p-k})(\bm{p}_\perp-\bm{k}_\perp)\cdot \bm{p}_\perp\\
        &=\frac{(eB)^2}{16\pi}\re^{-z}\frac{k!}{(l-1)!}(-z)^{l-k}L_{k}^{(l-k)}(z)L_{k-1}^{(l-k)}(z)\,,
    \end{split}
\end{equation}
\begin{equation}
\label{eq:Ckl}
    \begin{split}
        \Hfunc_{k,l}(z)&\coloneqq\int
        \frac{\dd[2]{p_\perp}}{(2\pi)^2}\re^{-2\xi_p}L^{(1)}_{k-1}(4\xi_p)
        \re^{-2\xi_{p-k}}L^{(1)}_{l-1}(4\xi_{p-k})\left[\bm{p}_\perp\cdot\bm{k}_\perp-\frac{2(\bm{p}_\perp\cdot\bm{k}_\perp)^2}{\bm{k}_\perp^2}+\bm{p}_\perp^2\right]\\ 
        &=\frac{(eB)^2}{16\pi}\re^{-z}\frac{k!}{(l-1)!}(-z)^{l-k}\left[\frac{k}{z}F(k,l,z)-\frac{k+l}{z}L_{k}^{(l-k)}(z)L_{k-1}^{(l-k)}(z)\right]\,,
    \end{split}
\end{equation}
\begin{equation}
\label{eq:Dkl}
    \begin{split}
        \Kfunc_{k,l}(z)&\coloneqq\int \frac{\dd[2]{p_\perp}}{(2\pi)^2}\re^{-2\xi_p}
        L^{(1)}_{k-1}(4\xi_p)\re^{-2\xi_{p-k}}
        L^{(1)}_{l-1}(4\xi_{p-k})\left[2\bm{p}_\perp^2-\frac{2(\bm{k}_\perp\cdot\bm{p}_\perp)^2}{\bm{k}_\perp^2}\right]\\
        &=\frac{(eB)^2}{16\pi}\re^{-z}\frac{k!}{(l-1)!}(-z)^{l-k}\left[
        \frac{k}{z}F(k,l,z)-\frac{k+l-z}{z}L_{k}^{(l-k)}(z)L_{k-1}^{(l-k)}(z)\right]\,,
    \end{split}
\end{equation}
and
\begin{equation}
\label{eq:Ekl}
\begin{split}
\mathcal{E}_{k,l}(z)&
\coloneqq\int \frac{\dd[2]{p_\perp}}{(2\pi)^2} \re^{-2\xi_p}
L^{(1)}_{k-1}(4\xi_p)
\re^{-2\xi_{p-k}}L_{l}(4\xi_{p-k})(\bm{p}_\perp\cdot\bm{k}_\perp)\\
&=2z\bigl[\Kfunc_{k,l}(z)+\Kfunc_{k,l+1}(z)\bigr]\,,
\end{split}
\end{equation}
where
\begin{align}
F(k,l,z)\coloneqq\Bigl[L_k^{(l-k)}(z)\Bigr]^2+\frac{l}{k}\Bigl[L_{k-1}^{(l-k)}(z)\Bigr]^2\,.
\end{align}
We use the same notation $F(k,l,z)$ as defined in ref.~\cite{Baier:2007dw}.
We can perform these integrations with the generating function for the generalized Laguerre polynomials (See Appendix~\ref{sec:transverse momentum integration}).
$\Ffunc_{k,l}(z),\Gfunc_{k,l}(z),\Hfunc_{k,l}(z)$, and $\Kfunc_{k,l}(z)$ are symmetric with respect to the replacement of $k$ and $l$ by definition, though their expressions seem not to be so.
To see this, the generalized Laguerre polynomials satisfy
\begin{equation}
    \frac{(-z)^k}{k!}L^{(k-l)}_{l}(z)=\frac{(-z)^l}{l!}L^{(l-k)}_{k}(z)\,,
\end{equation}
and using this property, we can easily check the symmetry with respect to $k$ and $l$.

It is now time to move on to the analytical calculation of $\tilde \LepT^{ij}_{kl}(p_\parallel, k)$.
Similarly to $\tilde \LepTdensity^{ij}_{kl}(p,k)$, $\tilde \LepT^{ij}_{kl}(p_\parallel,k)$ has only six independent components because of the Hermitian property.

For the diagonal components, after some calculations, we obtain

\begin{align}
\label{eq:h11}
&\tilLepT_{kl}^{11}(p_\parallel, k)\notag\\
%&=\frac{k^2}{k_\parallel^2}
%\biggl\{\bigl[2eB(k+l)-k_\parallel^2\bigr]
%\int\frac{\dd[2]{p_\perp}}{(2\pi)^2}\bigl[A_{k-}(\xi_p)A_{l+}(\xi_{p-k})+A_{k+}(\xi_p)A_{l-}(\xi_{p-k})\bigr]-\\
%&\quad-4\int\frac{\dd[2]{p_\perp}}{(2\pi)^2}\biggl[\bm{p}_\perp^2-\frac{2(\bm{p}_\perp\cdot \bm{k}_\perp)^2}{\bm{k}_\perp^2}+(\bm{p}_\perp\cdot \bm{k}_\perp) \biggr]B_k(\xi_p)B_l(\xi_{p-k})\biggl\}\\
&=\frac{k^2}{k_\parallel^2}
\Bigl\{\bigl[2eB(k+l)-k_\parallel^2\bigr]4(-1)^{l-k-1}\bigl[\Ffunc_{k-1,l}(z)+\Ffunc_{k,l-1}(z)\bigr]
-64(-1)^{l-k}\Hfunc_{k,l}(z)\Bigl\}\,,\\
%Similarly, the other components are
&\tilLepT_{kl}^{22}(p_\parallel, k)\notag\\
&=\bigl[2eB(k+l)-k_\parallel^2\bigr]4(-1)^{l-k-1}\bigl[\Ffunc_{k-1,l}(z)+\Ffunc_{k,l-1}(z)\bigr]+64(-1)^{l-k}\Hfunc_{k,l}(z)\,,\\
&\tilLepT_{kl}^{33}(p_\parallel, k)\notag\\
&=\frac{1}{k_\parallel^2}\Bigl\{\bigl[2eB(k-l)\bigr]^2
-k_\parallel^2\bigl[4m^2+2eB(k+l)\bigr]\Bigr\}4(-1)^{l-k}\bigl[\Ffunc_{k,l}(z)+\Ffunc_{k-1,l-1}(z)\bigr]-\notag\\
&\quad-64(-1)^{l-k}\Gfunc_{k,l}(z)\,.
\end{align}
Here, we used eqs.~\eqref{eq:h11_onshell}, \eqref{eq:h22_onshell}, and \eqref{eq:h33_onshell} for the integrands, $\tilLepTdensity_{kl}^{ij}(p_\parallel, k)$, simplified by the on-shell condition.

Similarly, the off diagonal components read
\begin{equation}
\label{eq:h12}
\tilLepT_{kl}^{12}(p_\parallel, k)=\frac{-\ri k^2}{\sqrt{k^2k_\parallel^2}}\bigl[2eB(k+l)-k_\parallel^2\bigr]4(-1)^{l-k-1}\bigl[\Ffunc_{k,l-1}(z)-\Ffunc_{k-1,l}(z)\bigr]\,,
\end{equation}
\begin{equation}
\begin{split}
\label{eq:h13}
\tilLepT_{kl}^{13}(p_\parallel, k)
&=\frac{- 2p_\parallel^\nu\epsilon_{\parallel\nu\lambda}k^\lambda}{k_\parallel^2}\frac{k^2}{\sqrt{\bm{k}_\perp^2k^2}}8(-1)^{l-k-1}
\bigl[\mathcal{E}_{k,l}(z)-\mathcal{E}_{k,l-1}(z)-\mathcal{E}_{l,k}(z)+\mathcal{E}_{l,k-1}(z)\bigr]\,,
\end{split}
\end{equation}
\begin{equation}
\begin{split}
\label{eq:h23}
\tilLepT_{kl}^{23}(p_\parallel, k)
&=\frac{-2\ri p_\parallel^\nu\epsilon_{\parallel \nu\lambda}k^\lambda}{ \sqrt{\bm{k}_\perp^2k_\parallel^2}}
8(-1)^{l-k-1}\bigl[
-\mathcal{E}_{k,l}(z)-\mathcal{E}_{k,l-1}(z)+\mathcal{E}_{l,k}(z)+\mathcal{E}_{l,k-1}(z)\bigr]\,.
\end{split}
\end{equation}
We used the on-shell condition for $\tilde \LepTdensity^{12}_{kl}(p_\parallel, k)$ only.
By contrast, $\tilde \LepT^{13}_{kl}(p_\parallel, k)$ and $\tilde \LepT^{23}_{kl}(p_\parallel, k)$ do not assume the on-shell condition.
With the on-shell condition, we find that $\tilde \LepT_{kl}^{13}(p_\parallel, k)$ and $\tilde \LepT_{kl}^{23}(p_\parallel, k)$ depend on the branch of the solution labeled by $t=\pm$, i.e.,
\begin{equation}
\begin{split}
&\tilLepT_{kl}^{13}(p_\parallel, k)\bigr|_{p_z=p_{zt}}\\
&=-tb_{kl}\frac{k^2}{\sqrt{\bm{k}_\perp^2k^2}}8(-1)^{l-k-1}
\bigl[\mathcal{E}_{k,l}(z)-\mathcal{E}_{k,l-1}(z)-\mathcal{E}_{l,k}(z)+\mathcal{E}_{l,k-1}(z)\bigr]\,,
\end{split}
\end{equation}
and
\begin{equation}
\begin{split}
&\tilLepT_{kl}^{23}(p_\parallel, k)\bigr|_{p_z=p_{zt}}\\
&=-\ri tb_{kl}\frac{k_\parallel^2}{ \sqrt{\bm{k}_\perp^2k_\parallel^2}}
8(-1)^{l-k-1}\bigl[
-\mathcal{E}_{k,l}(z)-\mathcal{E}_{k,l-1}(z)+\mathcal{E}_{l,k}(z)+\mathcal{E}_{l,k-1}(z)\bigr]\,.
\end{split}
\end{equation}

%%%%%%%%%%%%%%
\section{Numerical results}
Now we have all the analytical expressions necessary for the polarization tensor.
In this section, we numerically take the sum over the Landau levels with respect to $k$ and $l$, and numerically perform the principal value integration to get the real part of the polarization tensor.

We can decompose $\tilde{\Pi}_R^{ij}(k)$ to the vacuum part and the medium part as
\begin{equation}
\begin{split}
\tilde{\Pi}_R^{ij}(k) = \tilde{\Pi}_{R,\mathrm{vac}}^{ij}(k)+\tilde{\Pi}_{R,\mathrm{med}}^{ij}(k)\,,
\end{split}
\end{equation}
where $\tilde \Pi^{ij}_{R,\mathrm{vac}}(k)\coloneqq\lim_{T,\mu\rightarrow 0}\tilde \Pi^{ij}_{R}(k)$.
First, we shall discuss the vacuum part.
The imaginary part of the vacuum part reads
\begin{equation}
\begin{split}
2\Im\tilde{\Pi}_{R,\mathrm{vac}}^{ij}(k)
 &= -\sgn(k_0)\sum_{k,l=0}^\infty \theta\bigl(k_\parallel^2-(m_l+m_k)^2\bigr) \sum_{t=\pm}\frac{e^2\tilLepT_{kl}^{ij}(p_\parallel, k)\bigr|_{p_z=p_{zt}}}{2k_\parallel^2b_{kl}}\,.
\end{split}
\end{equation}
Only the diagonal components with $i=j$ take nonzero values because $\tilLepT_{kl}^{ij}(p_\parallel, k)|_{p_z=p_{zt}}$ with $i\neq j$ are antisymmetric with respect to $k$ and $l$, while $b_{kl}$ is symmetric.
In addition, the diagonal components are independent of $t$; see eqs.~\eqref{eq:h11_onshell}, \eqref{eq:h22_onshell}, and \eqref{eq:h33_onshell}.
After all, we find
\begin{equation}
\begin{split}
2\Im\tilde{\Pi}_{R,\mathrm{vac}}^{ij}(k)
 &= -\delta^{ij}\mathrm{sgn}(k_0)\sum_{k,l=0}^\infty \theta\bigl(k_\parallel^2-(m_l+m_k)^2\bigr)\frac{e^2\tilLepT_{kl}^{ii}(p_\parallel, k)\bigr|_{p_z=p_{z+}}}{k_\parallel^2b_{kl}}\,.
 \label{eq:im_vac}
\end{split}
\end{equation}
This is consistent with the results in ref.~\cite{Baier:2007dw}.

Next, from eq.~\eqref{eq:2imPi}, the medium part is expressed as
\begin{align}\label{eq:im_med}
&2\Im\tilde{\Pi}_{R,\mathrm{med}}^{ij}(k)\notag\\
 &= \sum_{k,l=0}^\infty  \sum_{t=\pm}\frac{e^2\tilLepT_{kl}^{ij}(p_\parallel, k)\bigr|_{p_z=p_{zt}}}{2k_\parallel^2b_{kl}}\Bigl\{\notag\\
& -\bigl[n_\mathrm{F}(E_{k,t}-\mu)-n_\mathrm{F}(E_{l,-t}-\mu)\bigr]
\theta\bigl((m_k-m_l)^2-k_\parallel^2\bigr)\bigl[\theta(k_\parallel^2)\theta\bigl(k_0(m_k-m_l)\bigr)
-\theta(tk_z)\theta(-k_\parallel^2)\bigr]+\notag\\
&+ \bigl[n_\mathrm{F}(E_{k,t}+\mu)-n_\mathrm{F}(E_{l,-t}+\mu)\bigr]
\theta\bigl((m_k-m_l)^2-k_\parallel^2\bigr)\bigl[\theta(k_\parallel^2)\theta\bigl(k_0(m_l-m_k)\bigr)
-\theta(tk_z)\theta(-k_\parallel^2)\bigr]+\notag\\
& + \bigl[n_\mathrm{F}(E_{k,t}-\mu)+n_\mathrm{F}(E_{l,-t}+\mu)\bigr]\theta(k_0)\theta\bigl(k_\parallel^2-(m_l+m_k)^2\bigr)-\notag\\
& - \bigl[n_\mathrm{F}(E_{k,t}+\mu)+n_\mathrm{F}(E_{l,-t}-\mu)\bigr]\theta(-k_0)\theta\bigl(k_\parallel^2-(m_l+m_k)^2\bigr)
\Bigr\}\,.
\end{align}

%%%%%%
\subsection{Real photon limit}
In this work, we limit our consideration to the process involving the real photons that satisfy the on-shell condition $k^2=0$.
In view of the expression $\tilde \LepT_{kl}^{ij}(p_\parallel, k)$ in eqs.~\eqref{eq:h11}-\eqref{eq:h23}, $\tilde \LepT_{kl}^{1i}(p_\parallel, k)$ are propotional to $k^2$.
Thus, $\tilde\Pi^{1i}_{R}(k)$ vanish for real photons.
This corresponds to the fact that massless vector bosons have no longitudinal component.

Then, we only need to focus on the following $2\times 2$ matrix:
\begin{align}
    \mqty(\tilde\Pi^{22}_{R} & \tilde\Pi^{23}_{R} \\ \tilde\Pi^{32}_{R} & \tilde\Pi^{33}_{R}) = \mqty(\Re\tilde\Pi^{22}_{R} & \Re\tilde\Pi^{23}_{R} \\ \Re\tilde\Pi^{32}_{R} & \Re\tilde\Pi^{33}_{R}) + \ri \mqty(\Im\tilde\Pi^{22}_{R} & \Im\tilde\Pi^{23}_{R} \\ \Im\tilde\Pi^{32}_{R} & \Im\tilde\Pi^{33}_{R})\,.
\end{align}

%%%%%%%%%
\subsection{Summation over the Landau levels}
We numerically take the sum over the Landau levels with respect to $k$ and $l$.
In principle, there is no upper limit for $k$ and $l$, but the contribution of Landau levels sufficiently larger than the energy scale under consideration should be negligible and can be censored out at a certain value when performing numerical calculations.
In our system, the energy scale is determined by temperature $T$, chemical potential $\mu$, and parallel components of photon momentum $k_\parallel$.
Thus, in the present work, we set the maximal value of the Landau levels as $k_{\mathrm{max}}=100$ that satisfies
\begin{equation}
\label{kmax}
    2|eB|k_{\mathrm{max}} > \mathrm{max}\,\bigl\{T^2,\mu^2,k_\parallel^2\bigr\}\,.
\end{equation}

\begin{figure}
\centering
\begin{minipage}{0.49\columnwidth}
    \centering
    \includegraphics[width=0.99\columnwidth]{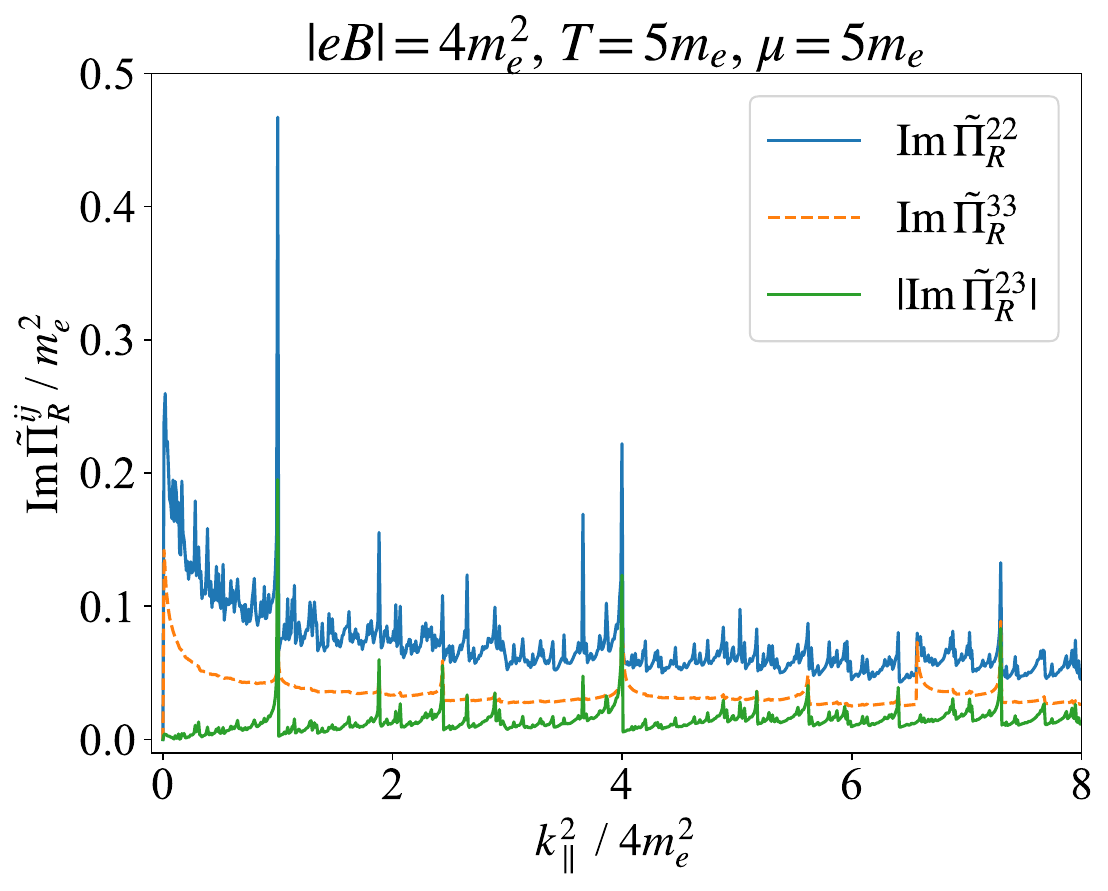}
\end{minipage}
\begin{minipage}{0.49\columnwidth}
    \centering
    \includegraphics[width=0.99\columnwidth]{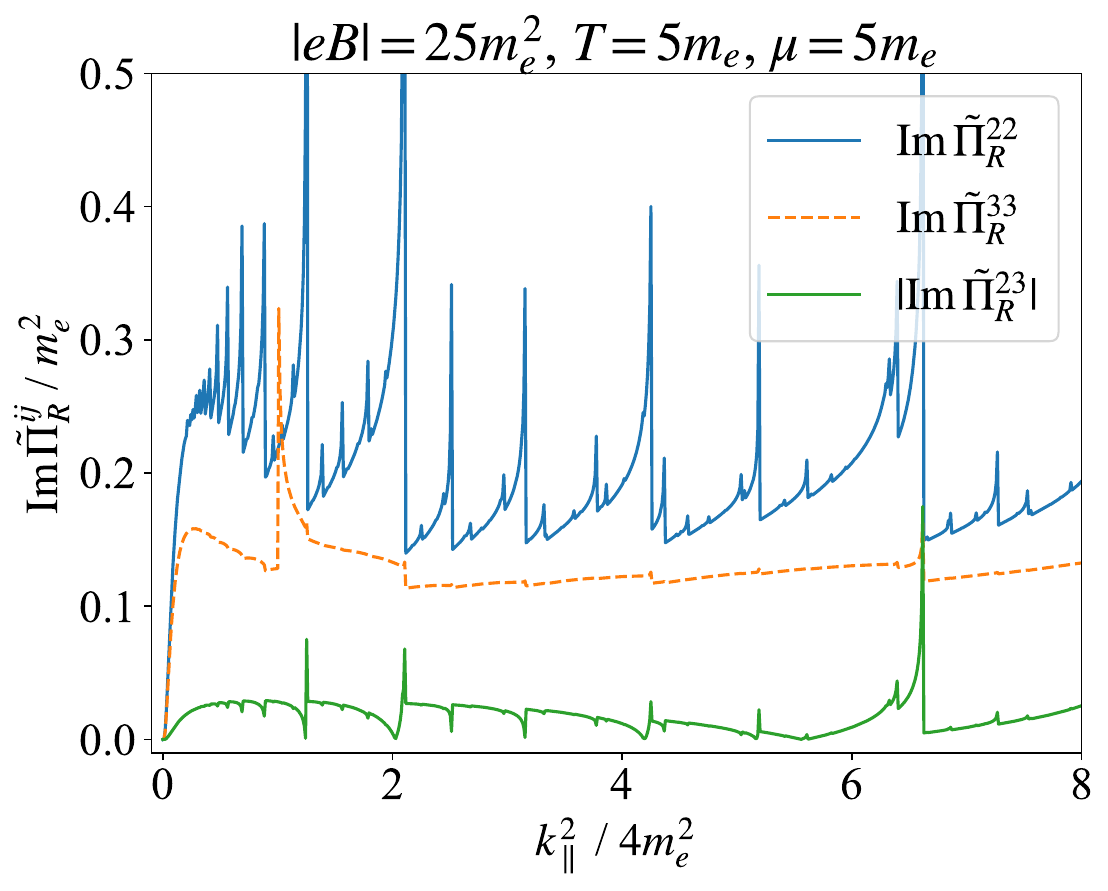}
\end{minipage}
\caption{Imaginary part of the polarization tensor components as functions of the photon momentum parallel to the magnetic field for $T=5m_e,\mu=5m_e$, and two fixed magnetic fields: $|eB| = 4m_e^2$ (left panel) and $|eB| = 25m_e^2$ (right panel).}
\label{fig:weak_field}
\end{figure}

We present the imaginary part of the polarization tensor after taking the Landau level sum in figure~\ref{fig:weak_field}.
In this figure, we show the result as a function of the photon momenta $k_\parallel^2$.
We note that the polarization function depends on two variables $k_0$ and $k_z$ independently.
Alternatively, we can introduce a parametrization of $k_z=k_0\cos\theta$ where $\theta$ is an angle between the photon momentum $\bm{k}$ and the magnetic field $\bm{B}$.
In this work, we present the results for the angle $\theta = \ang{45}$ only.

In figure~\ref{fig:weak_field}, the peaks in the imaginary part of the polarization tensor appear when the photon energy corresponds to the difference between the Landau levels.
We note that $\Im\tilde\Pi_{R}^{33}$ behaves more smoothly than $\Im\tilde\Pi_{R}^{22}$.
We recall that $\Im \tilde\Pi_{R}^{22}$ represents X-mode processes that require nonzero Landau level jumps, while O-mode processes in $\Im\tilde\Pi_{R}^{33}$ are allowed without Landau level jumps.
The intervals between the peaks are smaller for weaker $B$ (left panel), then they become larger for stronger $B$ (right panel).
This difference is understood from the gaps between the Landau levels which are enhanced by the magnetic field.
The off-diagonal components,  $\Im\tilde\Pi_R^{23}, \Im\tilde\Pi_R^{32}$, are actually pure imaginary. Nevertheless, we follow the convention to call the part calculated with the Cutkoski cutting rules the imaginary part. The off-diagonal components of the real part are also pure imaginary.
We note $|\Im\tilde\Pi_R^{23}|\sim\mathcal{O}(z^1)$ and $\Im\tilde\Pi_R^{22}, \Im\tilde\Pi_R^{33}\sim \mathcal{O}(z^0)$, where $z$ is suppressed for strong $B$.
Thus, we see that $|\Im\tilde\Pi_R^{23}|$ is relatively more suppressed than $\Im\tilde\Pi_R^{22}, \Im\tilde\Pi_R^{33}$ when $B$ is strong (right panel).

\begin{figure}
\centering
\begin{minipage}{0.49\columnwidth}
    \centering
    \includegraphics[width=0.99\columnwidth]{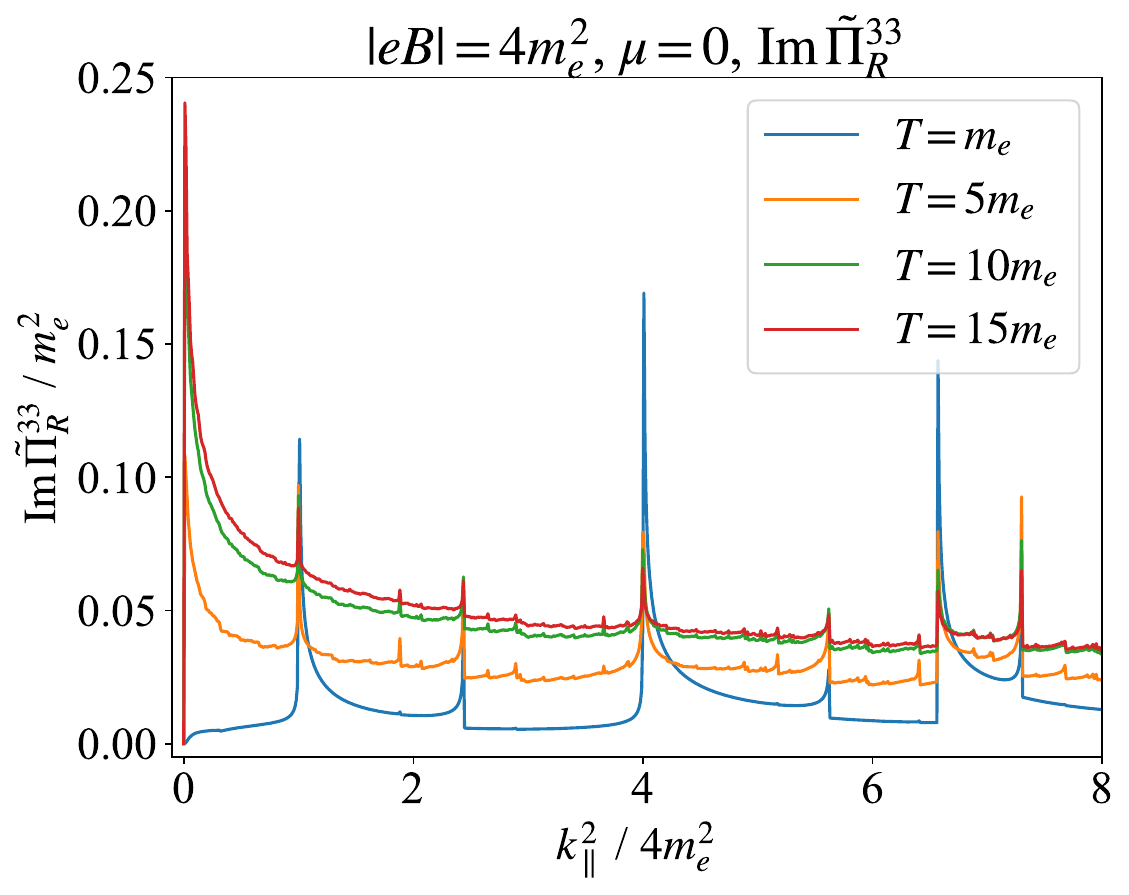}
\end{minipage}
\begin{minipage}{0.49\columnwidth}
    \centering
    \includegraphics[width=0.99\columnwidth]{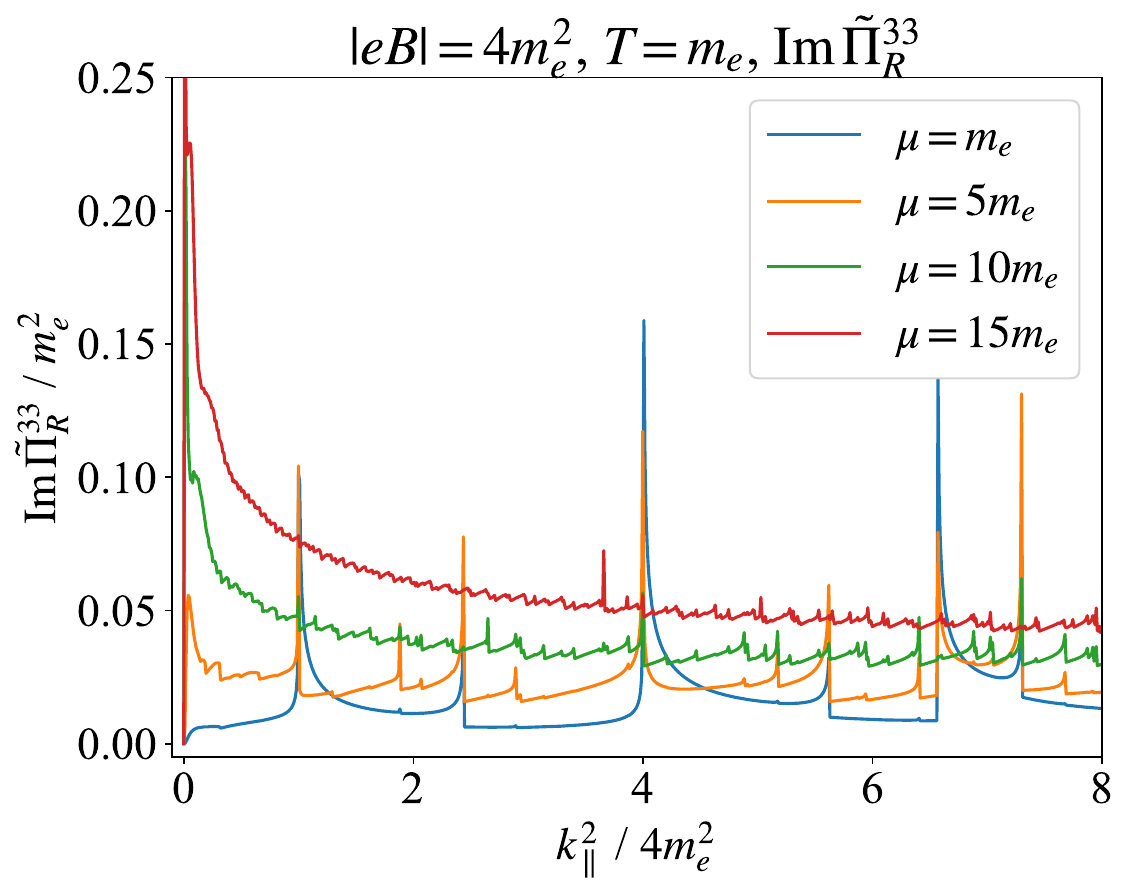}
\end{minipage}
\caption{Imaginary part of the $33$-component of the polarization tensor component as a function of the photon momentum parallel to the magnetic field at $|eB|=4m_e^2$.  (Left panel) Dependence on four different values of $T$ at $\mu=0$.  (Right panel) Dependence on four different values of $\mu$ at $T=m_e$.}
\label{fig:mu_dep}
\end{figure}

In addition, we present the $\Im\tilde\Pi_{R}^{33}$ for various temperatures and chemical potentials in figure~\ref{fig:mu_dep}.
At low temperature and low density, we can see spectra corresponding to photon splitting into a pair of electron and positron.
As the temperature or the chemical potential increases, the effect of absorption by the medium gets larger and the effect of photon splitting becomes relatively small.

%--- figure ---%
%--- figure ---%

%--- figure ---%
%--- figure ---%

%%%%%%%%%
\subsection{Photon decay rates}
\begin{figure}
    \centering
    \includegraphics[width=0.25\textwidth]{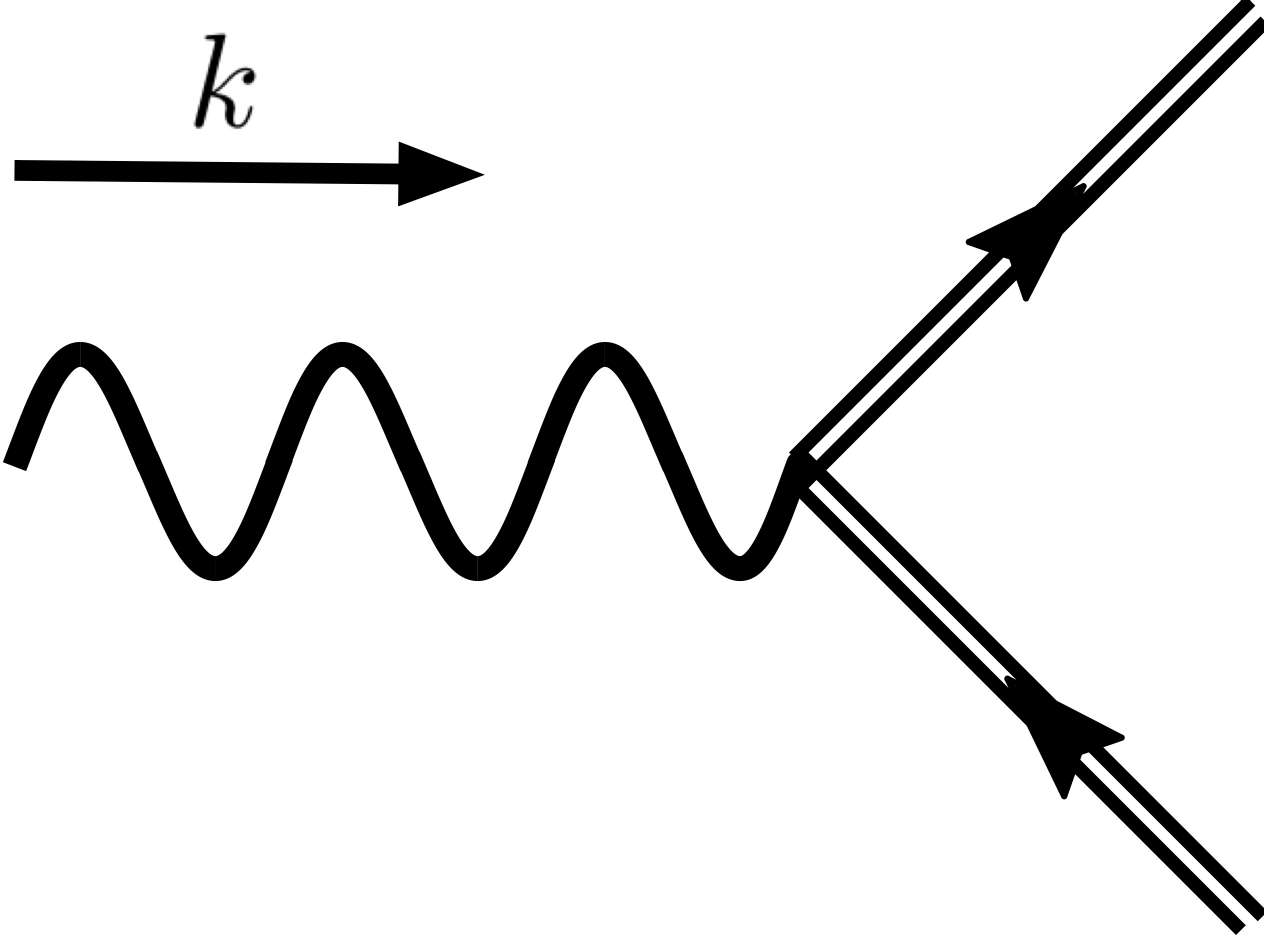}
    \caption{Photon splitting process into an electron and a positron}
    \label{fig:decay}
\end{figure}

\begin{figure}
\centering
\begin{minipage}{0.49\columnwidth}
    \centering
    \includegraphics[width=0.55\columnwidth]{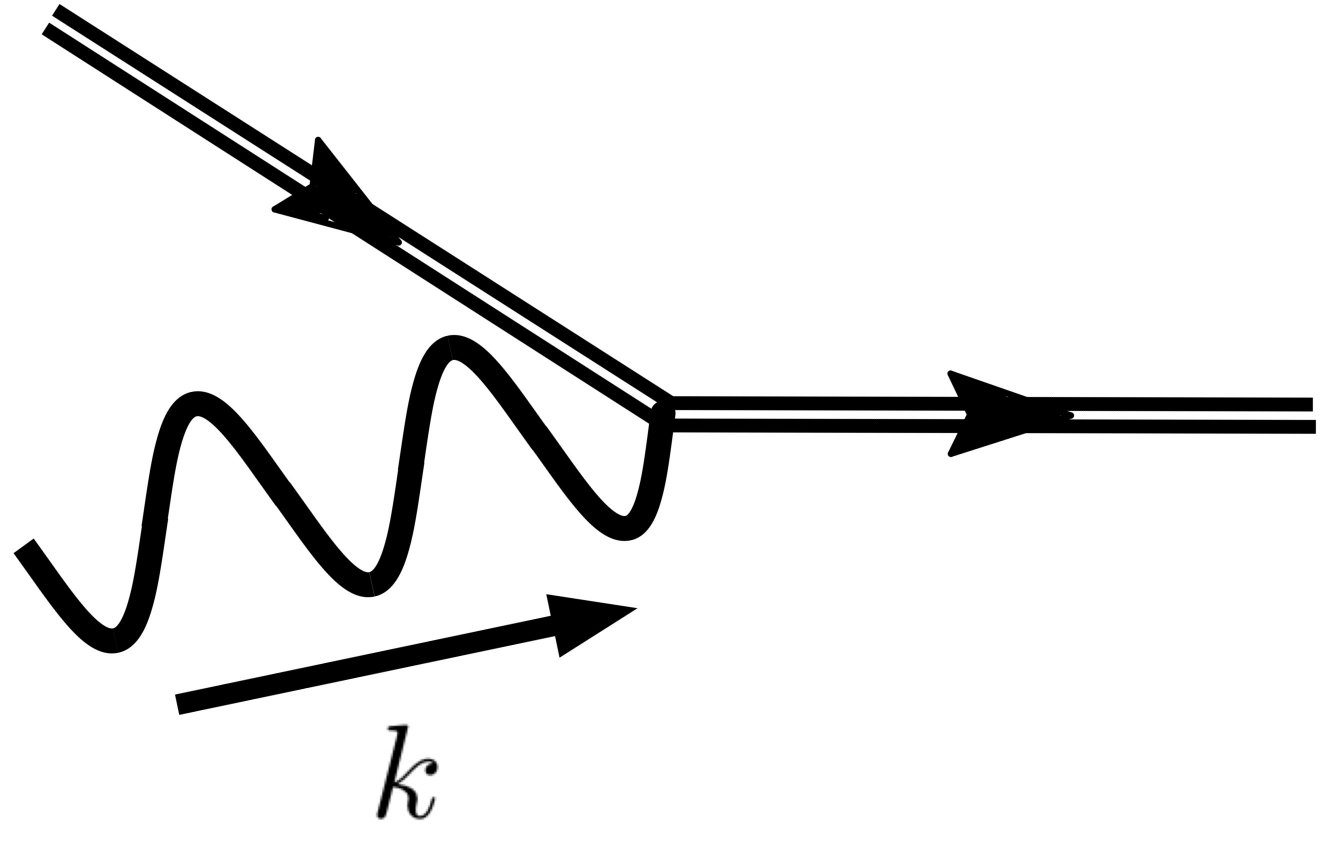}
\end{minipage}
\begin{minipage}{0.49\columnwidth}
    \centering
    \includegraphics[width=0.55\columnwidth]{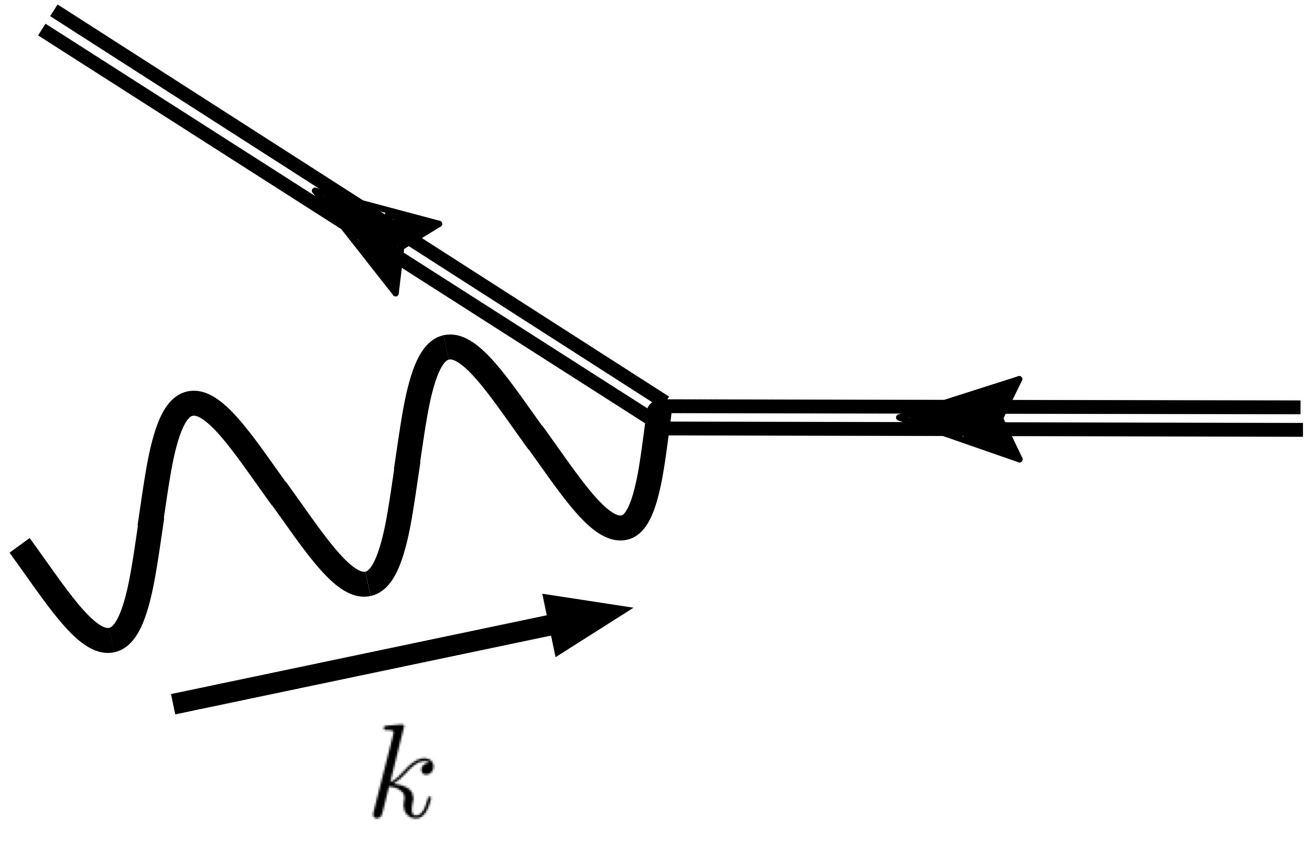}
\end{minipage}
\caption{Photon absorption processes by an electron (left panel) and by a positron (right panel).}
\label{fig:absorption}
\end{figure}

One of the physical quantities given in terms of the imaginary part of the photon polarization tensor is the photon decay rates in a hot and dense medium:
\begin{align}
\label{eq:rate}
    k_0\dv[3]{\Gamma^\mathrm{X}}{k}=\dfrac{1}{(2\pi)^3}\dfrac{\Im\tilde\Pi_R^{22}}{\re^{\beta k_0}-1}\,,\qquad k_0\dv[3]{\Gamma^\mathrm{O}}{k}=\dfrac{1}{(2\pi)^3}\dfrac{\Im\tilde\Pi_R^{33}}{\re^{\beta k_0}-1}\,.
\end{align}
The decay rates for $i=2$ and $3$ correspond to the X-mode photons and the O-mode photons, respectively.
We note that the polarization tensor involving $i = 1$ does not contribute to the decay rates in the real photon limit.
The formula contains the processes of the splitting into an electron and a positron in figure~\ref{fig:decay} and the absorption by an electron and a positron in the magnetized medium in figure~\ref{fig:absorption}.

\begin{figure}
\centering
\begin{minipage}{0.49\columnwidth}
    \centering
    \includegraphics[width=0.99\columnwidth]{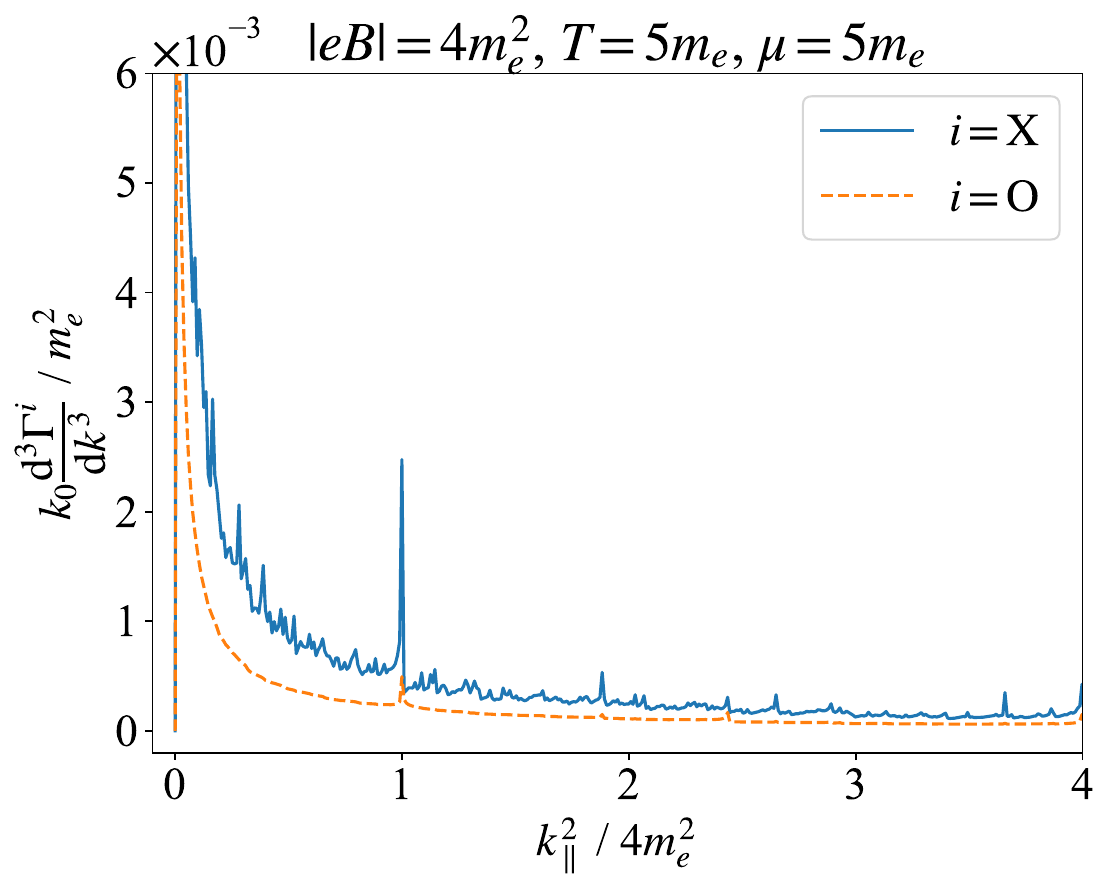}
\end{minipage}
\begin{minipage}{0.49\columnwidth}
    \centering
    \includegraphics[width=0.99\columnwidth]{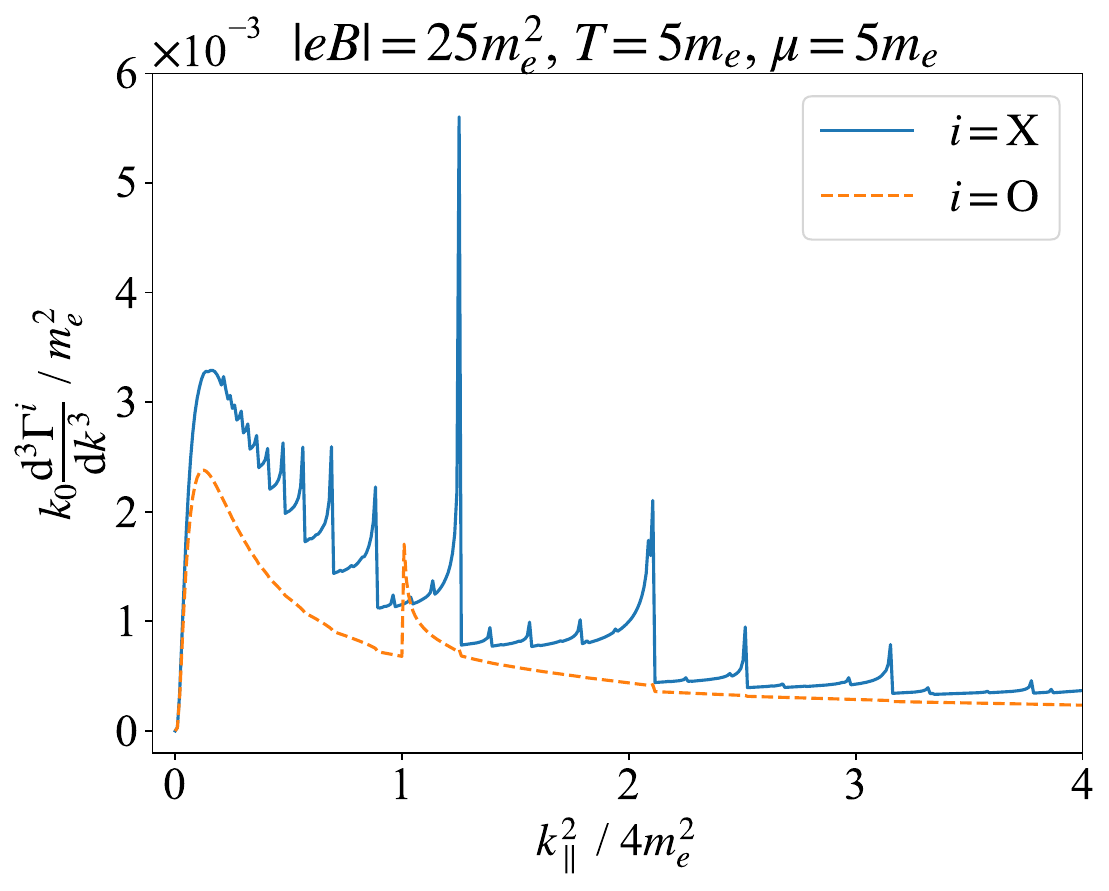}
\end{minipage}
\caption{The photon decay rates for two fixed magnetic field: $|eB| = 4m_e^2$ (left panel) and $|eB| = 25m_e^2$ (right panel).}
\label{fig:rate}
\end{figure}

The numerical results of these rates are shown in figure~\ref{fig:rate}. 
The rate decreases with increasing photon energy because of the smaller scattering amplitude of the excitation to the large Landau levels.
%and the Bose function in eq.~\eqref{eq:rate} suppresses the rates as the photon energy increases.
The peak at $k_\parallel^2 / (4m_e^2)=1$ in figure~\ref{fig:rate} corresponds to the onset of the splitting process of the photon into a pair of electron and positron.

\begin{figure}
    \centering
    \includegraphics[width=0.6\textwidth]{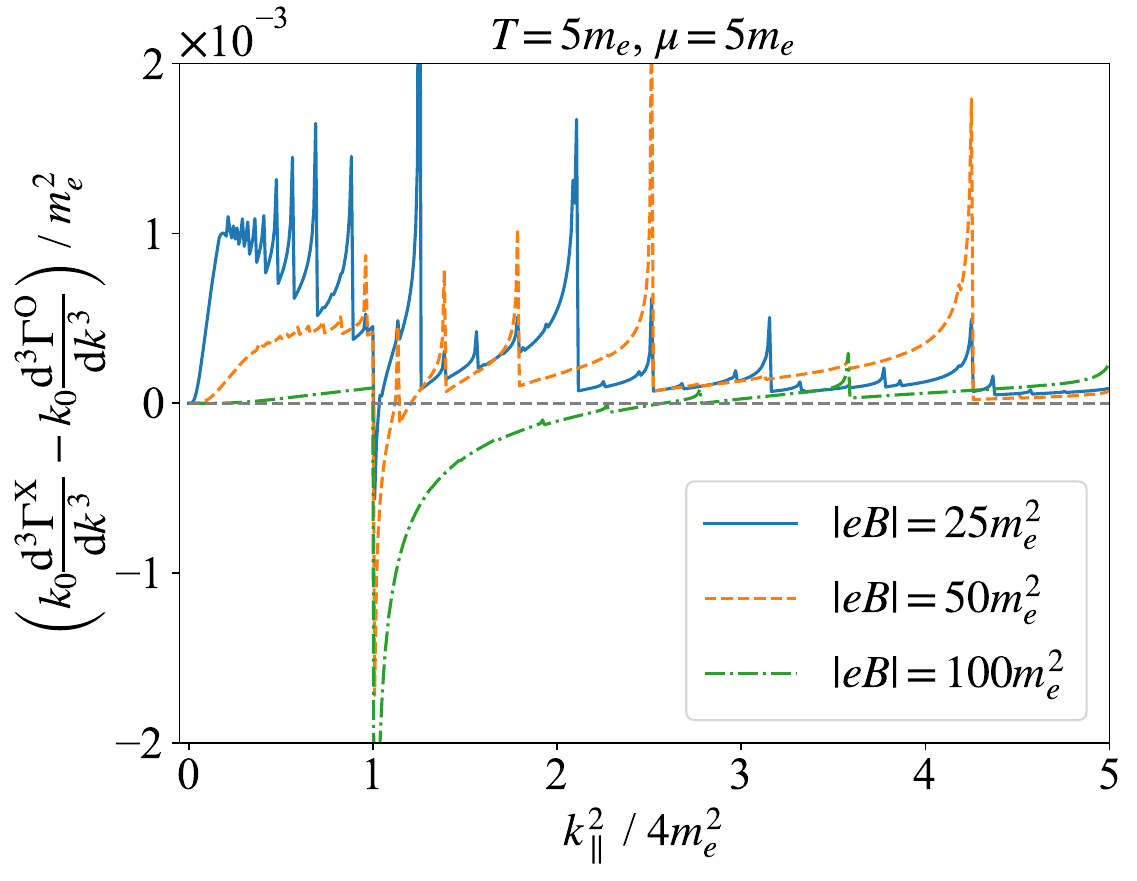}
    \caption{Difference between decay rates of two modes for three fixed magnetic fields: $|eB| = 25m_e^2$, $|eB| = 50m_e^2$, and $|eB| = 100m_e^2$.}
    \label{fig:rate_diff}
\end{figure}

For practical application, it is important to quantify how much X-mode photons or O-mode photons are more absorbed by the magnetized medium.
Thus, we calculate the difference in decay rates of the two modes, and the results are shown in figure~\ref{fig:rate_diff}.
The X-mode decay rate ($i=2$) is dominant over the O-mode decay rate ($i=3$) up to the moderate strength of the magnetic field $|eB| \simeq 50m_e^2$.
For the stronger magnetic field case, on the other hand, the O-mode overwhelms the decay rate.
For the former observation at a weak magnetic field, a semi-classical picture is sensible.  Charged particles are accelerated by $B$ which absorb and emit photons with the polarization along the plane of the particle motion orthogonal to the magnetic field.  Thus, such photons are X-mode dominated.  In contrast, for the strong magnetic field, we can understand the latter from the expected suppression of the X-mode associated with large Landau level gaps. 
 Which is more dominant in extreme environments is crucial for the account of observed polarized photons~\cite{Lai:2022knd}.

%--- figure ---%
%--- figure ---%

%--- figure ---%
%--- figure ---%

%--- figure ---%
%--- figure ---%

%--- figure ---%
%--- figure ---%

%%%%%%%%%%
\subsection{Principal value integral for the real part}
\begin{figure}
    \centering
    \includegraphics[width=0.6\textwidth]{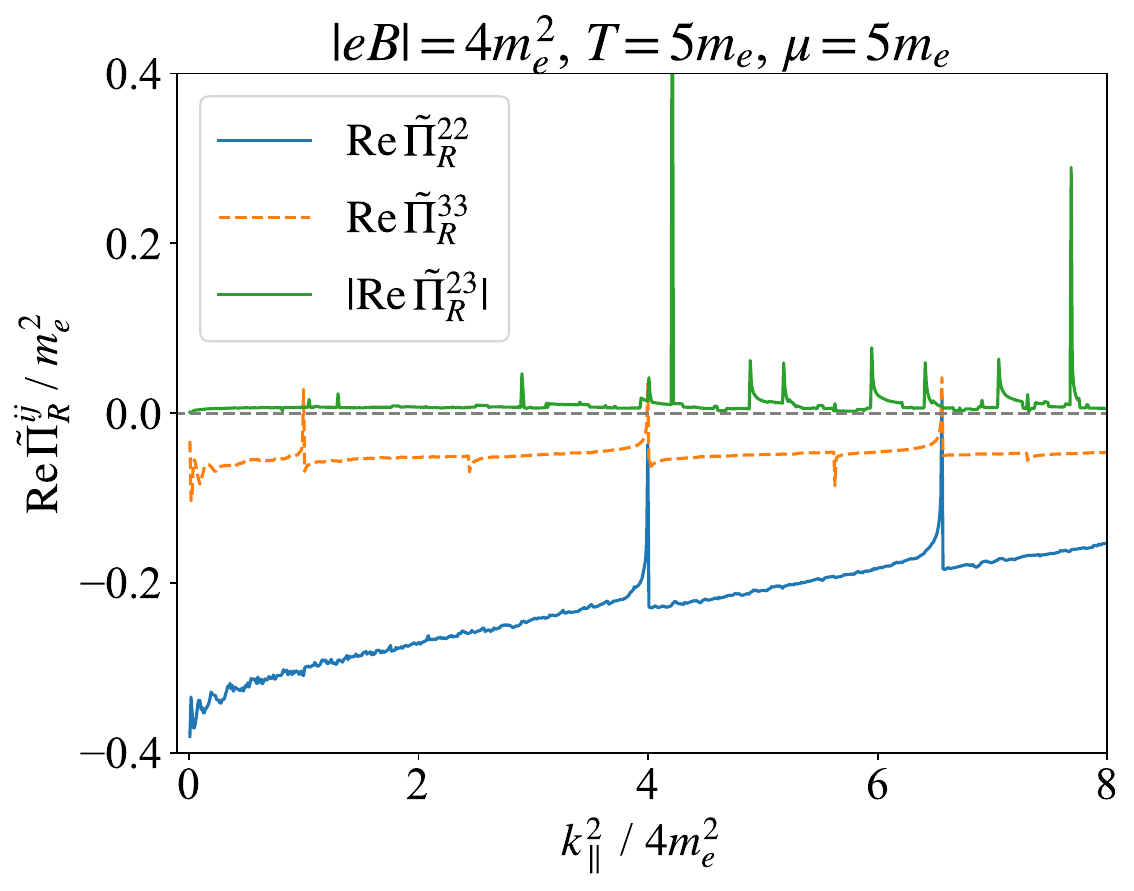}
    \caption{Real part of the polarization tensor as a function of the photon momentum parallel to the magnetic field for $T=5m_e,\mu=5m_e, |eB|=4m_e^2$.}
    \label{fig:real}
\end{figure}

To calculate the real part of the polarization tensor $\Re \tilde\Pi_{R}^{ij}(k)$, we numerically perform the principal value integral~\eqref{eq:Kramers-Kronig}.
We show the real part in figure~\ref{fig:real}.
In general, we have to deal with the ultraviolet divergence in the vacuum part.
However, the ultraviolet divergence vanishes for $k^2=0$, so that we can safely conduct the integration; see discussions in ref.~\cite{Hattori:2012je}.

Once we calculate both the real and imaginary parts of the polarization tensor, we can determine the polarization state of the eigenmodes.
The photon polarization in a magnetized medium is focused on, especially in the context of the polarized X-ray observation from magnetar sources~\cite{Lai:2022knd}.
The literature indicates that the polarization state changes from X(O)-mode to O(X)-mode with varying plasma density by taking into account the vacuum birefringence and the plasma oscillation.
In the next section, we will demonstrate this behavior on the basis of the polarization tensor at finite density, although our calculation is for an electron-positron system rather than an ion-electron system.

%--- figure ---%
%--- figure ---%

%%%%%%%%%%%
\subsection{Stokes parameters}
To see the behavior of the polarization state in a magnetized medium, we calculate the polarization vector, $\bm{E}=(E_1, E_2, E_3)$, by solving Maxwell's equations:
\begin{equation}\label{eq:Maxwell}
    \left(k^2\delta^{ij} + \tilde\Pi_{R}^{ij}\right)E_j = 0\,,\qquad (i, j=1,2,3)\,.
\end{equation}
When we take the real photon limit, $\tilde\Pi_R^{1i}$ vanishes, and we can ignore $E_1$.
Here, we redefine $E_\text{X}\coloneqq E_2$ and $E_\text{O}\coloneqq E_3$ to make the physical meaning easy to understand.
We can express the eigenvectors of the eq.~\eqref{eq:Maxwell} as a Jones vector:
\begin{equation}
    \bm{E}_{\pm} = \mqty(E_{\text{X}\pm} \\ E_{\text{O}\pm}) = \mqty(A_{\text{X}\pm}\re^{\ri\phi_{\text{X}\pm}} \\ A_{\text{O}\pm}\re^{\ri\phi_{\text{O}\pm}})\,,
\end{equation}
where $\pm$ denotes the two eigenmodes corresponding to each of the two eigenvalues.
$A_{\text{X}\pm}$ and $A_{\text{O}\pm}$ are amplitude, and $\phi_{\text{X}\pm}$ and $\phi_{\text{O}\pm}$ are phase.
Using this Jones vector, we can define Stokes parameters, which specify the polarization states of the photon, as
\begin{align}
    I_\pm&\coloneqq |E_{\text{X}\pm}|^2 + |E_{\text{O}\pm}|^2\,,\\
    Q_\pm&\coloneqq |E_{\text{X}\pm}|^2 - |E_{\text{O}\pm}|^2\,,\\
    U_\pm&\coloneqq E_{\text{X}\pm}E_{\text{O}\pm}^* + E_{\text{X}\pm}^*E_{\text{O}\pm}\,,\\
    V_\pm&\coloneqq \ri(E_{\text{X}\pm}E_{\text{O}\pm}^* - E_{\text{X}\pm}^*E_{\text{O}\pm})\,.
\end{align}

\begin{figure}
\centering
\begin{minipage}{0.49\columnwidth}
    \centering
    \includegraphics[width=0.99\columnwidth]{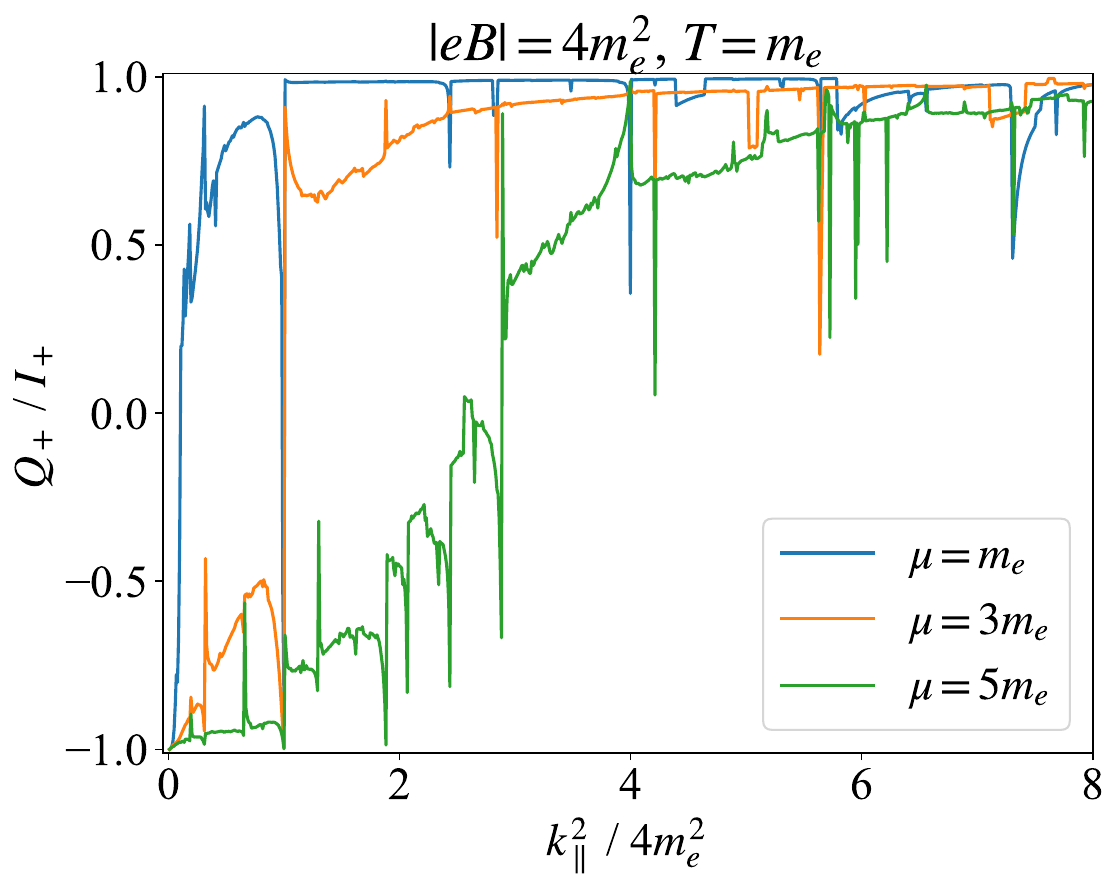}
\end{minipage}
\begin{minipage}{0.49\columnwidth}
    \centering
    \includegraphics[width=0.99\columnwidth]{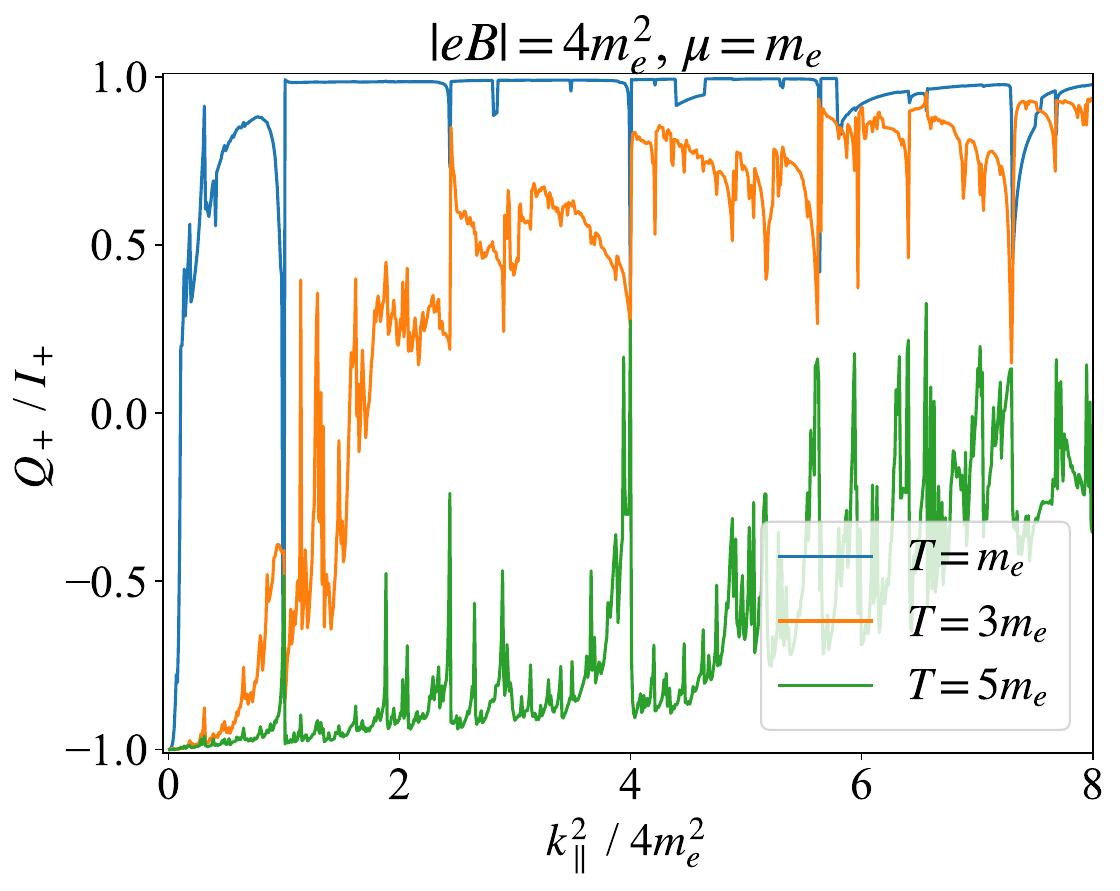}
\end{minipage}
\caption{The normalized Stokes parameters, $Q_+/I_+$, as functions of photon momenta.
In the left panel, temperature is fixed at $T=m_e$, and chemical potential takes three values.
In the right panel, chemical potential is fixed at $\mu=m_e$, and temperature takes three values.}
\label{fig:Q}
\end{figure}

Since our interest is in which mode the photon is polarized in a magnetized medium, we calculate $Q_\pm/I_\pm$.
Here, $I_\pm$ is the absolute value of the Jones vector, and then $Q_\pm/I_\pm$ is normalized.
$Q_\pm$ represents a difference in the intensity between the two modes.
If $Q_\pm/I_\pm$ is $+1$, the eigenmode is polarized in X-mode.
If $Q_\pm/I_\pm$ is $-1$, on the other hand, the eigenmode is polarized in O-mode.
We show the calculation result as functions of photon momenta in figure~\ref{fig:Q}.
In this calculation, we use $\bm{E}_+$ as the eigenmode, and we show $Q_+/I_+$.
$Q_+$ and $Q_-$ differ only in sign because the two eigenmodes are orthogonal to each other.

From this figure, we can see that the polarization state can change as photon energy, temperature, and chemical potential change.
This behavior is unique to finite density.
When the chemical potential is zero, the eigenmode is linearly polarized in one specific direction due to the absence of the off-diagonal components of the polarization tensor.

Focusing specifically on the low momenta region in the left panel of the figure~\ref{fig:Q}, we can see $\bm{E}_+$ is polarized in X-mode at low density but polarized in O-mode at high density.
Although it looks like the photon propagating in the environment with varying density changes its polarization state, it may convert into the other eigenmode around $Q_\pm = 0$, so that the polarization state may not change.
In other words, if the photon in the $\bm{E}_+$-eigenmode proceeds from the high-density region to the low-density region, the polarization state changes from O-mode to X-mode. On the other hand, if the photon jumps from $\bm{E}_+$ to $\bm{E}_-$-eigenmode around $Q_\pm=0$, it stays O-mode during the propagation.
% This behavior is called "mode conversion", and it is indicated in ref.~\cite{Lai:2022knd} by taking into account the vacuum birefringence and the plasma oscillation.
% The literature points to the importance of the mode conversion as the explanation for the polarized X-ray observation from magnetar sources.
% Although our calculation is for an electron-positron system rather than an ion-electron system, this is the first demonstration of mode conversion from a field theoretical calculation of finite temperature and density.

%--- figure ---%
%--- figure ---%

%%%%%%%%%
\section{Conclusions}
In this paper, we calculated the photon polarization tensor, $\Pi^{\mu\nu}$, at finite temperature and density in a constant magnetic field.
Since the magnetic field breaks rotational symmetry in the system, the structure of the photon polarization tensor becomes complicated.
We performed the tensor decomposition and mapped the two transverse components of the polarization tensor to the components parallel and orthogonal to the magnetic field, respectively.
For the fermion loop in the polarization tensor, we analytically took the Matsubara sum and imposed the energy conservation constraint according to the calculation procedures in the preceding works.  We found closed formulas for the integration with respect to transverse momenta, which correctly reproduce the equivalent results obtained in a different approach~\cite{Wang:2021eud}.
As a result, we obtained the expressions for $\Im \Pi^{\mu\nu}$ which is to be summed numerically over the Landau levels.
Specifically, we focused on the real photons with the on-shell condition and summed up the Landau levels for $\Im \Pi^{\mu\nu}$ with truncation up to the first one hundred levels.
We checked the justification of truncation by monitoring that the energy above the truncated level is sufficiently larger than the relevant energy scales of the temperature, the density, and the photon momenta.
We then discussed the temperature and the density dependence of $\Im \Pi^{\mu\nu}$ which is generally enhanced as the medium effects increase.

The calculation results show how the polarization tensor is sensitive to the polarization modes.
For an application example, we evaluated the decay rate of photons with different polarizations in the hot and dense magnetized medium.
In particular, the decay rate of photons orthogonal to the magnetic field, i.e., X-mode photons, is larger than that of photons parallel to the magnetic field, i.e., O-mode photons, as long as the magnetic field is not strong.
We also confirmed that the X-mode is suppressed by the Landau level gaps at a strong magnetic field, and then the O-mode dominates the photon decay rate.

We then proceed to our original calculations to estimate the real part of the polarization tensor.
With resulting $\Im \Pi^{\mu\nu}$, we utilized the Kramers-Kronig relation and calculated $\Re \Pi^{\mu\nu}$ numerically.
Since we focused on the real photons only, no ultraviolet divergence arises in the calculation of the vacuum part of $\Re \Pi^{\mu\nu}$, and the renormalization procedure is unnecessary.
From $\Re \Pi^{\mu\nu}$, we calculated the Stokes parameter which characterizes the polarization state.
We demonstrated that the O-mode dominant state is converted to the X-mode dominant state, and vice versa, as the electron density changes.
Although there is a quantitative difference between the electron system and the electron-ion system, our results of the polarization conversion with increasing density are qualitatively consistent with the astrophysical scenario to explain the origin of the polarization change observed by the IXPE~\cite{Lai:2022knd}.

The polarization dependence we have clarified is expected to be applied to phenomenological analyses on electromagnetic probes from hot and dense magnetized QCD matter created in the relativistic heavy-ion collision and polarized photons from the atmosphere at the surface of the magnetar.
For a more complete picture of the phenomenological application, especially for the polarized photons from the magnetar, we should add the ion contributions, and we are making progress along these lines.

\acknowledgments
The authors thank
Dong~Lai and
Toru~Tamagawa
for discussions.
TU was supported by FoPM, WINGS Program, the University of Tokyo.
%This work was supported by JSPS KAKENHI Grant Numbers~
This work was supported by Japan Society for the Promotion of Science
(JSPS) KAKENHI Grant Nos.\
22H01216, 22H05118 (KF), and
21H01084, 24H00975 (YH).

\appendix
%%%%%%%%%%%
\section{Formulas for the transverse momentum integration}\label{sec:transverse momentum integration}
In this section, we prove the formulas in eqs.~\eqref{eq:Akl}-\eqref{eq:Ekl}.

At first, we compute the following integral:
\begin{equation}
    \begin{split}
        \Ffunc_{k,l}(z)&=\int \frac{\dd[2]{p_\perp}}{(2\pi)^2} \re^{-2\xi_p}L_{k}(4\xi_p)
        \re^{-2\xi_{p-k}}L_{l}(4\xi_{p-k})\,.
    \end{split}
\end{equation}
In order to calculate this, it is useful to remind oneself of the generating functions for the generalized Laguerre polynomials:
\begin{equation}
    \sum_{n=0}^\infty t^nL_n^{(\alpha)}(x)=\dfrac{\re^{-tx/(1-t)}}{(1-t)^{\alpha+1}}\,.
\end{equation}
Using this, we define the generating function for $\Ffunc_{k,l}(z)$ as
\begin{equation}
\begin{split}
\label{eq:generate_A}
\Ffunc(s,t,z)\coloneqq&\sum_{k=0}^\infty\sum_{l=0}^\infty s^kt^l\Ffunc_{k,l}(z)\\
=&\frac{1}{1-s}\frac{1}{1-t}\int \frac{\dd[2]{p_\perp}}{(2\pi)^2} \re^{-2\xi_p}
\re^{-2\xi_{p-k}}
\re^{-\frac{4s}{1-s}\xi_p}
\re^{-\frac{4t}{1-t}\xi_{p-k}}\\
=&\frac{1}{1-s}\frac{1}{1-t}\int \frac{\dd[2]{p_\perp}}{(2\pi)^2}
\re^{-\frac{1+s}{1-s}\frac{\bm{p}_\perp^2}{eB}}
\re^{-\frac{1+t}{1-t}\frac{\bm{p}_\perp^2+\bm{k}_\perp^2-2\bm{p}_\perp\cdot\bm{k}_\perp}{eB}}\,.
\end{split}
\end{equation}
Replacing $\alpha=(1+s)/(1-s)$, and $\beta=(1+t)/(1-t)$,
we obtain
\begin{equation}
\begin{split}
\Ffunc(s,t,z)&=\frac{1}{1-s}\frac{1}{1-t}\int \frac{\dd[2]{p_\perp}}{(2\pi)^2}
\re^{-\frac{\alpha}{eB}\bm{p}_\perp^2-\frac{\beta}{eB}(\bm{p}_\perp^2+\bm{k}_\perp^2-2\bm{p}_\perp\cdot \bm{k}_\perp)}\\
&=\frac{1}{1-s}\frac{1}{1-t}\int \frac{\dd[2]{p_\perp}}{(2\pi)^2}
\re^{-\frac{\alpha+\beta}{eB}\left(\bm{p}_\perp-\frac{\beta}{\alpha+\beta}\bm{k}_\perp\right)^2-\frac{\alpha\beta}{eB(\alpha+\beta)}\bm{k}_\perp^2}\,.
\end{split}
\end{equation}
Conducting Gaussian integration, we find
\begin{equation}
\begin{split}
\label{eq:genereate_A_final}
\Ffunc(s,t,z)&=\frac{1}{1-s}\frac{1}{1-t}\frac{1}{4\pi}\frac{eB}{\alpha+\beta}
\re^{-\frac{\alpha\beta}{eB(\alpha+\beta)}\bm{k}_\perp^2}\\
&=\frac{eB}{8\pi}\re^{-z}\frac{1}{1-st}
\re^{-\frac{s+t+2st}{1-st}z}\\
&=\frac{eB}{8\pi}\re^{-z}
\sum_{k=0}^{\infty}\sum_{l=0}^{\infty}s^kt^l\frac{k!}{l!}(-z)^{l-k}\Bigl[L_{k}^{(l-k)}(z)\Bigr]^2\,.
\end{split}
\end{equation}
Comparing eq.~\eqref{eq:genereate_A_final} with eq.~\eqref{eq:generate_A} yields the final result:
\begin{equation}
    \begin{split}
        \Ffunc_{k,l}(z)&=\frac{eB}{8\pi}\re^{-z}\frac{k!}{l!}(-z)^{l-k}\Bigl[L_{k}^{(l-k)}(z)\Bigr]^2\,.
    \end{split}
\end{equation}

Similarly, the generating functions for other quantities can be introduced as
\begin{equation}
\begin{split}
\Gfunc(s,t,z)&\coloneqq
\frac{s}{(1-s)^2}\frac{t}{(1-t)^2}\int \frac{\dd[2]{p_\perp}}{(2\pi)^2}
\re^{-\frac{\bm{p}_\perp^2}{eB}}
\re^{-\frac{(\bm{p}_\perp-\bm{k}_\perp)^2}{eB}}
\re^{-\frac{2s}{1-s}\frac{\bm{p}_\perp^2}{eB}}
\re^{-\frac{2t}{1-t}\frac{(\bm{p}_\perp-\bm{k}_\perp)^2}{eB}}\times\\
&\qquad\times \left(\bm{p}_\perp-\bm{k}_\perp\right)\cdot\bm{p}_\perp\\
&=\frac{(eB)^2}{16\pi}
\frac{st}{(1-st)^2}
\left[1-\frac{(1+s)(1+t)}{1-st}z\right]
\re^{-\frac{(1+s)(1+t)}{1-st}z}\\
&=\frac{(eB)^2}{16\pi}\frac{st}{(1-st)^2}
\left(1+z\frac{d}{dz}
\right)
\re^{-\frac{(1+s)(1+t)}{1-st}z}\\
&=
\frac{(eB)^2}{16\pi}\re^{-z}\sum_{k=1}^\infty\sum_{l=1}^\infty s^{k}t^{l}\frac{k!}{(l-1)!} (-z)^{l-k}
L_{k}^{(l-k)}(z)L_{k-1}^{(l-k)}(z)\,,
\end{split}
\end{equation}
\begin{equation}
\begin{split}
\Hfunc(s,t,z)&\coloneqq
\frac{s}{(1-s)^2}\frac{t}{(1-t)^2}\int \frac{\dd[2]{p_\perp}}{(2\pi)^2}
\re^{-\frac{\bm{p}_\perp^2}{eB}}
\re^{-\frac{(\bm{p}_\perp-\bm{k}_\perp)^2}{eB}}
\re^{-\frac{2s}{1-s}\frac{\bm{p}_\perp^2}{eB}}
\re^{-\frac{2t}{1-t}\frac{(\bm{p}_\perp-\bm{k}_\perp)^2}{eB}}\times\\
&\qquad
\times\left[\bm{p}_\perp\cdot\bm{k}_\perp-\frac{2(\bm{p}_\perp\cdot\bm{k}_\perp)^2}{\bm{k}_\perp^2}+\bm{p}_\perp^2\right]\\
&=\frac{(eB)^2}{16\pi}\frac{st(1+s)(1+t)}{(1-st)^3}z\re^{-\frac{(1+s)(1+t)}{1-st}z}\\
&=\frac{(eB)^2}{16\pi}\re^{-z}\sum_{k=1}^\infty\sum_{l=1}^\infty s^{k}t^{l}
\frac{k!}{(l-1)!}(-z)^{l-k}\times\\
&\qquad\times \left[\frac{k}{z}F(k,l,z)-\frac{k+l}{z}L_{k}^{(l-k)}(z)L_{k-1}^{(l-k)}(z)\right]\,,
\end{split}
\end{equation}

\begin{equation}
\begin{split}
\Kfunc(s,t,z)&\coloneqq
\frac{s}{(1-s)^2}\frac{t}{(1-t)^2}\int \frac{\dd[2]{p_\perp}}{(2\pi)^2}
\re^{-\frac{\bm{p}_\perp^2}{eB}}
\re^{-\frac{(\bm{p}_\perp-\bm{k}_\perp)^2}{eB}}
\re^{-\frac{2s}{1-s}\frac{\bm{p}_\perp^2}{eB}}
\re^{-\frac{2t}{1-t}\frac{(\bm{p}_\perp-\bm{k}_\perp)^2}{eB}}\times\\
&\qquad
\times\left[2\bm{p}_\perp^2-\frac{2(\bm{p}_\perp\cdot\bm{k}_\perp)^2}{\bm{k}_\perp^2}\right]\\
&=\frac{(eB)^2}{16\pi}\frac{st}{(1-st)^2}\re^{-\frac{(1+s)(1+t)}{1-st}z}\\
&=\frac{(eB)^2}{16\pi}\re^{-z}\sum_{k=1}^\infty\sum_{l=1}^\infty s^{k}t^{l}
\frac{k!}{(l-1)!}(-z)^{l-k}\times\\
&\qquad\times \biggl[\frac{k}{z}F(k,l,z)-\frac{k+l-z}{z}L_{k}^{(l-k)}(z)L_{k-1}^{(l-k)}(z)\biggr]\,,
\end{split}
\end{equation}
and
\begin{equation}
\begin{split}
\mathcal{E}(s,t,z)&\coloneqq
\frac{s}{(1-s)^2}\frac{1}{1-t}\int \frac{\dd[2]{p_\perp}}{(2\pi)^2}
\re^{-\frac{\bm{p}_\perp^2}{eB}}
\re^{-\frac{(\bm{p}_\perp-\bm{k}_\perp)^2}{eB}}
\re^{-\frac{2s}{1-s}\frac{\bm{p}_\perp^2}{eB}}
\re^{-\frac{2t}{1-t}\frac{(\bm{p}_\perp-\bm{k}_\perp)^2}{eB}}\left(\bm{p}_\perp\cdot\bm{k}_\perp\right)\\
&=\frac{(eB)^2}{16\pi}\frac{2s(1+t)}{(1-st)^2}z\re^{-\frac{(1+s)(1+t)}{1-st}z}\\
&=2z\biggl[\frac{1}{t}\Kfunc(s,t,z)+\Kfunc(s,t,z)\biggr]\\
&=\sum_{k=1}^{\infty}\sum_{l=0}^{\infty}s^{k}t^{l}2z\bigl[D_{k,l+1}(z)+D_{k,l}(z)\bigr]
\,.
\end{split}
\end{equation}
With these generating functions, the formulas of our interest are immediately recovered.

\bibliographystyle{JHEP}
\bibliography{photon_polarization}

\providecommand{\href}[2]{#2}\begingroup\raggedright\begin{thebibliography}{10}

\bibitem{Fraga:2008qn}
E.S.~Fraga and A.J.~Mizher, \emph{{Chiral transition in a strong magnetic
  background}}, \href{https://doi.org/10.1103/PhysRevD.78.025016}{\emph{Phys.
  Rev. D} {\bfseries 78} (2008) 025016}
  [\href{https://arxiv.org/abs/0804.1452}{{\ttfamily 0804.1452}}].

\bibitem{Mizher:2010zb}
A.J.~Mizher, M.N.~Chernodub and E.S.~Fraga, \emph{{Phase diagram of hot QCD in
  an external magnetic field: possible splitting of deconfinement and chiral
  transitions}}, \href{https://doi.org/10.1103/PhysRevD.82.105016}{\emph{Phys.
  Rev. D} {\bfseries 82} (2010) 105016}
  [\href{https://arxiv.org/abs/1004.2712}{{\ttfamily 1004.2712}}].

\bibitem{Ferrer:2013noa}
E.J.~Ferrer, V.~de~la Incera, I.~Portillo and M.~Quiroz, \emph{{New look at the
  QCD ground state in a magnetic field}},
  \href{https://doi.org/10.1103/PhysRevD.89.085034}{\emph{Phys. Rev. D}
  {\bfseries 89} (2014) 085034}
  [\href{https://arxiv.org/abs/1311.3400}{{\ttfamily 1311.3400}}].

\bibitem{Gatto:2010qs}
R.~Gatto and M.~Ruggieri, \emph{{Dressed Polyakov loop and phase diagram of hot
  quark matter under magnetic field}},
  \href{https://doi.org/10.1103/PhysRevD.82.054027}{\emph{Phys. Rev. D}
  {\bfseries 82} (2010) 054027}
  [\href{https://arxiv.org/abs/1007.0790}{{\ttfamily 1007.0790}}].

\bibitem{Gatto:2010pt}
R.~Gatto and M.~Ruggieri, \emph{{Deconfinement and Chiral Symmetry Restoration
  in a Strong Magnetic Background}},
  \href{https://doi.org/10.1103/PhysRevD.83.034016}{\emph{Phys. Rev. D}
  {\bfseries 83} (2011) 034016}
  [\href{https://arxiv.org/abs/1012.1291}{{\ttfamily 1012.1291}}].

\bibitem{Ferreira:2014kpa}
M.~Ferreira, P.~Costa, O.~Louren\c{c}o, T.~Frederico and C.~Provid\^encia,
  \emph{{Inverse magnetic catalysis in the (2+1)-flavor Nambu-Jona-Lasinio and
  Polyakov-Nambu-Jona-Lasinio models}},
  \href{https://doi.org/10.1103/PhysRevD.89.116011}{\emph{Phys. Rev. D}
  {\bfseries 89} (2014) 116011}
  [\href{https://arxiv.org/abs/1404.5577}{{\ttfamily 1404.5577}}].

\bibitem{Mueller:2015fka}
N.~Mueller and J.M.~Pawlowski, \emph{{Magnetic catalysis and inverse magnetic
  catalysis in QCD}},
  \href{https://doi.org/10.1103/PhysRevD.91.116010}{\emph{Phys. Rev. D}
  {\bfseries 91} (2015) 116010}
  [\href{https://arxiv.org/abs/1502.08011}{{\ttfamily 1502.08011}}].

\bibitem{Klimenko:1991he}
K.G.~Klimenko, \emph{{Three-dimensional Gross-Neveu model at nonzero
  temperature and in an external magnetic field}},
  \href{https://doi.org/10.1007/BF01566663}{\emph{Z. Phys. C} {\bfseries 54}
  (1992) 323}.

\bibitem{Gusynin:1994re}
V.P.~Gusynin, V.A.~Miransky and I.A.~Shovkovy, \emph{{Catalysis of dynamical
  flavor symmetry breaking by a magnetic field in (2+1)-dimensions}},
  \href{https://doi.org/10.1103/PhysRevLett.73.3499}{\emph{Phys. Rev. Lett.}
  {\bfseries 73} (1994) 3499}
  [\href{https://arxiv.org/abs/hep-ph/9405262}{{\ttfamily hep-ph/9405262}}].

\bibitem{Gusynin:1995nb}
V.P.~Gusynin, V.A.~Miransky and I.A.~Shovkovy, \emph{{Dimensional reduction and
  catalysis of dynamical symmetry breaking by a magnetic field}},
  \href{https://doi.org/10.1016/0550-3213(96)00021-1}{\emph{Nucl. Phys. B}
  {\bfseries 462} (1996) 249}
  [\href{https://arxiv.org/abs/hep-ph/9509320}{{\ttfamily hep-ph/9509320}}].

\bibitem{Shushpanov:1997sf}
I.A.~Shushpanov and A.V.~Smilga, \emph{{Quark condensate in a magnetic field}},
  \href{https://doi.org/10.1016/S0370-2693(97)00441-3}{\emph{Phys. Lett. B}
  {\bfseries 402} (1997) 351}
  [\href{https://arxiv.org/abs/hep-ph/9703201}{{\ttfamily hep-ph/9703201}}].

\bibitem{Fukushima:2012xw}
K.~Fukushima and J.M.~Pawlowski, \emph{{Magnetic catalysis in hot and dense
  quark matter and quantum fluctuations}},
  \href{https://doi.org/10.1103/PhysRevD.86.076013}{\emph{Phys. Rev. D}
  {\bfseries 86} (2012) 076013}
  [\href{https://arxiv.org/abs/1203.4330}{{\ttfamily 1203.4330}}].

\bibitem{Bali:2013txa}
G.S.~Bali, F.~Bruckmann, G.~Endr\"odi and A.~Sch\"afer, \emph{{Magnetization
  and pressures at nonzero magnetic fields in QCD}},
  \href{https://doi.org/10.22323/1.187.0182}{\emph{PoS} {\bfseries LATTICE2013}
  (2014) 182} [\href{https://arxiv.org/abs/1310.8145}{{\ttfamily 1310.8145}}].

\bibitem{Miransky:2015ava}
V.A.~Miransky and I.A.~Shovkovy, \emph{{Quantum field theory in a magnetic
  field: From quantum chromodynamics to graphene and Dirac semimetals}},
  \href{https://doi.org/10.1016/j.physrep.2015.02.003}{\emph{Phys. Rept.}
  {\bfseries 576} (2015) 1} [\href{https://arxiv.org/abs/1503.00732}{{\ttfamily
  1503.00732}}].

\bibitem{DElia:2010abb}
M.~D'Elia, S.~Mukherjee and F.~Sanfilippo, \emph{{QCD Phase Transition in a
  Strong Magnetic Background}},
  \href{https://doi.org/10.1103/PhysRevD.82.051501}{\emph{Phys. Rev. D}
  {\bfseries 82} (2010) 051501}
  [\href{https://arxiv.org/abs/1005.5365}{{\ttfamily 1005.5365}}].

\bibitem{Bali:2011qj}
G.S.~Bali, F.~Bruckmann, G.~Endrodi, Z.~Fodor, S.D.~Katz, S.~Krieg et~al.,
  \emph{{The QCD phase diagram for external magnetic fields}},
  \href{https://doi.org/10.1007/JHEP02(2012)044}{\emph{JHEP} {\bfseries 02}
  (2012) 044} [\href{https://arxiv.org/abs/1111.4956}{{\ttfamily 1111.4956}}].

\bibitem{Endrodi:2024cqn}
G.~Endrodi, \emph{{QCD with background electromagnetic fields on the lattice: a
  review}},  \href{https://arxiv.org/abs/2406.19780}{{\ttfamily 2406.19780}}.

\bibitem{Bali:2012zg}
G.S.~Bali, F.~Bruckmann, G.~Endrodi, Z.~Fodor, S.D.~Katz and A.~Schafer,
  \emph{{QCD quark condensate in external magnetic fields}},
  \href{https://doi.org/10.1103/PhysRevD.86.071502}{\emph{Phys. Rev. D}
  {\bfseries 86} (2012) 071502}
  [\href{https://arxiv.org/abs/1206.4205}{{\ttfamily 1206.4205}}].

\bibitem{Bruckmann:2013oba}
F.~Bruckmann, G.~Endrodi and T.G.~Kovacs, \emph{{Inverse magnetic catalysis and
  the Polyakov loop}},
  \href{https://doi.org/10.1007/JHEP04(2013)112}{\emph{JHEP} {\bfseries 04}
  (2013) 112} [\href{https://arxiv.org/abs/1303.3972}{{\ttfamily 1303.3972}}].

\bibitem{Hidaka:2012mz}
Y.~Hidaka and A.~Yamamoto, \emph{{Charged vector mesons in a strong magnetic
  field}}, \href{https://doi.org/10.1103/PhysRevD.87.094502}{\emph{Phys. Rev.
  D} {\bfseries 87} (2013) 094502}
  [\href{https://arxiv.org/abs/1209.0007}{{\ttfamily 1209.0007}}].

\bibitem{Hattori:2015aki}
K.~Hattori, T.~Kojo and N.~Su, \emph{{Mesons in strong magnetic fields: (I)
  General analyses}},
  \href{https://doi.org/10.1016/j.nuclphysa.2016.03.016}{\emph{Nucl. Phys. A}
  {\bfseries 951} (2016) 1} [\href{https://arxiv.org/abs/1512.07361}{{\ttfamily
  1512.07361}}].

\bibitem{Ding:2020jui}
H.-T.~Ding, S.-T.~Li, S.~Mukherjee, A.~Tomiya and X.-D.~Wang, \emph{{Meson
  masses in external magnetic fields with HISQ fermions}},
  \href{https://doi.org/10.22323/1.363.0250}{\emph{PoS} {\bfseries LATTICE2019}
  (2020) 250} [\href{https://arxiv.org/abs/2001.05322}{{\ttfamily
  2001.05322}}].

\bibitem{Skokov:2009qp}
V.~Skokov, A.Y.~Illarionov and V.~Toneev, \emph{{Estimate of the magnetic field
  strength in heavy-ion collisions}},
  \href{https://doi.org/10.1142/S0217751X09047570}{\emph{Int. J. Mod. Phys. A}
  {\bfseries 24} (2009) 5925}
  [\href{https://arxiv.org/abs/0907.1396}{{\ttfamily 0907.1396}}].

\bibitem{Voronyuk:2011jd}
V.~Voronyuk, V.D.~Toneev, W.~Cassing, E.L.~Bratkovskaya, V.P.~Konchakovski and
  S.A.~Voloshin, \emph{{(Electro-)Magnetic field evolution in relativistic
  heavy-ion collisions}},
  \href{https://doi.org/10.1103/PhysRevC.83.054911}{\emph{Phys. Rev. C}
  {\bfseries 83} (2011) 054911}
  [\href{https://arxiv.org/abs/1103.4239}{{\ttfamily 1103.4239}}].

\bibitem{Deng:2012pc}
W.-T.~Deng and X.-G.~Huang, \emph{{Event-by-event generation of electromagnetic
  fields in heavy-ion collisions}},
  \href{https://doi.org/10.1103/PhysRevC.85.044907}{\emph{Phys. Rev. C}
  {\bfseries 85} (2012) 044907}
  [\href{https://arxiv.org/abs/1201.5108}{{\ttfamily 1201.5108}}].

\bibitem{McLerran:2013hla}
L.~McLerran and V.~Skokov, \emph{{Comments About the Electromagnetic Field in
  Heavy-Ion Collisions}},
  \href{https://doi.org/10.1016/j.nuclphysa.2014.05.008}{\emph{Nucl. Phys. A}
  {\bfseries 929} (2014) 184}
  [\href{https://arxiv.org/abs/1305.0774}{{\ttfamily 1305.0774}}].

\bibitem{Kharzeev:2007jp}
D.E.~Kharzeev, L.D.~McLerran and H.J.~Warringa, \emph{{The Effects of
  topological charge change in heavy ion collisions: 'Event by event P and CP
  violation'}},
  \href{https://doi.org/10.1016/j.nuclphysa.2008.02.298}{\emph{Nucl. Phys. A}
  {\bfseries 803} (2008) 227}
  [\href{https://arxiv.org/abs/0711.0950}{{\ttfamily 0711.0950}}].

\bibitem{Fukushima:2008xe}
K.~Fukushima, D.E.~Kharzeev and H.J.~Warringa, \emph{{The Chiral Magnetic
  Effect}}, \href{https://doi.org/10.1103/PhysRevD.78.074033}{\emph{Phys. Rev.
  D} {\bfseries 78} (2008) 074033}
  [\href{https://arxiv.org/abs/0808.3382}{{\ttfamily 0808.3382}}].

\bibitem{Son:2004tq}
D.T.~Son and A.R.~Zhitnitsky, \emph{{Quantum anomalies in dense matter}},
  \href{https://doi.org/10.1103/PhysRevD.70.074018}{\emph{Phys. Rev. D}
  {\bfseries 70} (2004) 074018}
  [\href{https://arxiv.org/abs/hep-ph/0405216}{{\ttfamily hep-ph/0405216}}].

\bibitem{Metlitski:2005pr}
M.A.~Metlitski and A.R.~Zhitnitsky, \emph{{Anomalous axion interactions and
  topological currents in dense matter}},
  \href{https://doi.org/10.1103/PhysRevD.72.045011}{\emph{Phys. Rev. D}
  {\bfseries 72} (2005) 045011}
  [\href{https://arxiv.org/abs/hep-ph/0505072}{{\ttfamily hep-ph/0505072}}].

\bibitem{Son:2009tf}
D.T.~Son and P.~Surowka, \emph{{Hydrodynamics with Triangle Anomalies}},
  \href{https://doi.org/10.1103/PhysRevLett.103.191601}{\emph{Phys. Rev. Lett.}
  {\bfseries 103} (2009) 191601}
  [\href{https://arxiv.org/abs/0906.5044}{{\ttfamily 0906.5044}}].

\bibitem{Kharzeev:2010gr}
D.E.~Kharzeev and D.T.~Son, \emph{{Testing the chiral magnetic and chiral
  vortical effects in heavy ion collisions}},
  \href{https://doi.org/10.1103/PhysRevLett.106.062301}{\emph{Phys. Rev. Lett.}
  {\bfseries 106} (2011) 062301}
  [\href{https://arxiv.org/abs/1010.0038}{{\ttfamily 1010.0038}}].

\bibitem{Kharzeev:2004ey}
D.~Kharzeev, \emph{{Parity violation in hot QCD: Why it can happen, and how to
  look for it}},
  \href{https://doi.org/10.1016/j.physletb.2005.11.075}{\emph{Phys. Lett. B}
  {\bfseries 633} (2006) 260}
  [\href{https://arxiv.org/abs/hep-ph/0406125}{{\ttfamily hep-ph/0406125}}].

\bibitem{Kharzeev:2013ffa}
D.E.~Kharzeev, \emph{{The Chiral Magnetic Effect and Anomaly-Induced
  Transport}}, \href{https://doi.org/10.1016/j.ppnp.2014.01.002}{\emph{Prog.
  Part. Nucl. Phys.} {\bfseries 75} (2014) 133}
  [\href{https://arxiv.org/abs/1312.3348}{{\ttfamily 1312.3348}}].

\bibitem{Shi:2017cpu}
S.~Shi, Y.~Jiang, E.~Lilleskov and J.~Liao, \emph{{Anomalous Chiral Transport
  in Heavy Ion Collisions from Anomalous-Viscous Fluid Dynamics}},
  \href{https://doi.org/10.1016/j.aop.2018.04.026}{\emph{Annals Phys.}
  {\bfseries 394} (2018) 50}
  [\href{https://arxiv.org/abs/1711.02496}{{\ttfamily 1711.02496}}].

\bibitem{Kharzeev:2015znc}
D.E.~Kharzeev, J.~Liao, S.A.~Voloshin and G.~Wang, \emph{{Chiral magnetic and
  vortical effects in high-energy nuclear collisions\textemdash{}A status
  report}}, \href{https://doi.org/10.1016/j.ppnp.2016.01.001}{\emph{Prog. Part.
  Nucl. Phys.} {\bfseries 88} (2016) 1}
  [\href{https://arxiv.org/abs/1511.04050}{{\ttfamily 1511.04050}}].

\bibitem{STAR:2017ckg}
{\scshape STAR} collaboration, \emph{{Global $\Lambda$ hyperon polarization in
  nuclear collisions: evidence for the most vortical fluid}},
  \href{https://doi.org/10.1038/nature23004}{\emph{Nature} {\bfseries 548}
  (2017) 62} [\href{https://arxiv.org/abs/1701.06657}{{\ttfamily 1701.06657}}].

\bibitem{Liang:2004ph}
Z.-T.~Liang and X.-N.~Wang, \emph{{Globally polarized quark-gluon plasma in
  non-central A+A collisions}},
  \href{https://doi.org/10.1103/PhysRevLett.94.102301}{\emph{Phys. Rev. Lett.}
  {\bfseries 94} (2005) 102301}
  [\href{https://arxiv.org/abs/nucl-th/0410079}{{\ttfamily nucl-th/0410079}}].

\bibitem{Becattini:2016gvu}
F.~Becattini, I.~Karpenko, M.~Lisa, I.~Upsal and S.~Voloshin, \emph{{Global
  hyperon polarization at local thermodynamic equilibrium with vorticity,
  magnetic field and feed-down}},
  \href{https://doi.org/10.1103/PhysRevC.95.054902}{\emph{Phys. Rev. C}
  {\bfseries 95} (2017) 054902}
  [\href{https://arxiv.org/abs/1610.02506}{{\ttfamily 1610.02506}}].

\bibitem{Alford:1999pb}
M.G.~Alford, J.~Berges and K.~Rajagopal, \emph{{Magnetic fields within color
  superconducting neutron star cores}},
  \href{https://doi.org/10.1016/S0550-3213(99)00830-5}{\emph{Nucl. Phys. B}
  {\bfseries 571} (2000) 269}
  [\href{https://arxiv.org/abs/hep-ph/9910254}{{\ttfamily hep-ph/9910254}}].

\bibitem{Ferrer:2005vd}
E.J.~Ferrer, V.~de~la Incera and C.~Manuel, \emph{{Magnetic color flavor
  locking phase in high density QCD}},
  \href{https://doi.org/10.1103/PhysRevLett.95.152002}{\emph{Phys. Rev. Lett.}
  {\bfseries 95} (2005) 152002}
  [\href{https://arxiv.org/abs/hep-ph/0503162}{{\ttfamily hep-ph/0503162}}].

\bibitem{Ferrer:2006vw}
E.J.~Ferrer, V.~de~la Incera and C.~Manuel, \emph{{Color-superconducting gap in
  the presence of a magnetic field}},
  \href{https://doi.org/10.1016/j.nuclphysb.2006.04.013}{\emph{Nucl. Phys. B}
  {\bfseries 747} (2006) 88}
  [\href{https://arxiv.org/abs/hep-ph/0603233}{{\ttfamily hep-ph/0603233}}].

\bibitem{Noronha:2007wg}
J.L.~Noronha and I.A.~Shovkovy, \emph{{Color-flavor locked superconductor in a
  magnetic field}},
  \href{https://doi.org/10.1103/PhysRevD.76.105030}{\emph{Phys. Rev. D}
  {\bfseries 76} (2007) 105030}
  [\href{https://arxiv.org/abs/0708.0307}{{\ttfamily 0708.0307}}].

\bibitem{Fukushima:2007fc}
K.~Fukushima and H.J.~Warringa, \emph{{Color superconducting matter in a
  magnetic field}},
  \href{https://doi.org/10.1103/PhysRevLett.100.032007}{\emph{Phys. Rev. Lett.}
  {\bfseries 100} (2008) 032007}
  [\href{https://arxiv.org/abs/0707.3785}{{\ttfamily 0707.3785}}].

\bibitem{Brauner:2016pko}
T.~Brauner and N.~Yamamoto, \emph{{Chiral Soliton Lattice and Charged Pion
  Condensation in Strong Magnetic Fields}},
  \href{https://doi.org/10.1007/JHEP04(2017)132}{\emph{JHEP} {\bfseries 04}
  (2017) 132} [\href{https://arxiv.org/abs/1609.05213}{{\ttfamily
  1609.05213}}].

\bibitem{Chen:2021vou}
S.~Chen, K.~Fukushima and Z.~Qiu, \emph{{Skyrmions in a magnetic field and
  \ensuremath{\pi}0 domain wall formation in dense nuclear matter}},
  \href{https://doi.org/10.1103/PhysRevD.105.L011502}{\emph{Phys. Rev. D}
  {\bfseries 105} (2022) L011502}
  [\href{https://arxiv.org/abs/2104.11482}{{\ttfamily 2104.11482}}].

\bibitem{Qiu:2023guy}
Z.~Qiu and M.~Nitta, \emph{{Quasicrystals in QCD}},
  \href{https://doi.org/10.1007/JHEP05(2023)170}{\emph{JHEP} {\bfseries 05}
  (2023) 170} [\href{https://arxiv.org/abs/2304.05089}{{\ttfamily
  2304.05089}}].

\bibitem{Evans:2023hms}
G.W.~Evans and A.~Schmitt, \emph{{Chiral Soliton Lattice turns into 3D
  crystal}}, \href{https://doi.org/10.1007/JHEP02(2024)041}{\emph{JHEP}
  {\bfseries 2024} (2024) 041}
  [\href{https://arxiv.org/abs/2311.03880}{{\ttfamily 2311.03880}}].

\bibitem{Taverna:2020vpr}
R.~Taverna, R.~Turolla, V.~Suleimanov, A.Y.~Potekhin and S.~Zane, \emph{{X-ray
  spectra and polarization from magnetar candidates}},
  \href{https://doi.org/10.1093/mnras/staa204}{\emph{Mon. Not. Roy. Astron.
  Soc.} {\bfseries 492} (2020) 5057}
  [\href{https://arxiv.org/abs/2001.07663}{{\ttfamily 2001.07663}}].

\bibitem{Taverna:2022jgl}
R.~Taverna et~al., \emph{{Polarized x-rays from a magnetar}},
  \href{https://arxiv.org/abs/2205.08898}{{\ttfamily 2205.08898}}.

\bibitem{Taverna:2024uop}
R.~Taverna and R.~Turolla, \emph{{X-ray Polarization from Magnetar Sources}},
  \href{https://doi.org/10.3390/galaxies12010006}{\emph{Galaxies} {\bfseries
  12} (2024) 6} [\href{https://arxiv.org/abs/2402.05622}{{\ttfamily
  2402.05622}}].

\bibitem{Zane:2023khc}
S.~Zane et~al., \emph{{A Strong X-Ray Polarization Signal from the Magnetar
  1RXS J170849.0-400910}},
  \href{https://doi.org/10.3847/2041-8213/acb703}{\emph{Astrophys. J. Lett.}
  {\bfseries 944} (2023) L27}
  [\href{https://arxiv.org/abs/2301.12919}{{\ttfamily 2301.12919}}].

\bibitem{Turolla:2023ruu}
R.~Turolla et~al., \emph{{IXPE and XMM-Newton Observations of the Soft Gamma
  Repeater SGR 1806\textendash{}20}},
  \href{https://doi.org/10.3847/1538-4357/aced05}{\emph{Astrophys. J.}
  {\bfseries 954} (2023) 88}
  [\href{https://arxiv.org/abs/2308.01238}{{\ttfamily 2308.01238}}].

\bibitem{Lai:2003nd}
D.~Lai and W.C.G.~Ho, \emph{{Polarized x-ray emission from magnetized neutron
  stars: Signature of strong - field vacuum polarization}},
  \href{https://doi.org/10.1103/PhysRevLett.91.071101}{\emph{Phys. Rev. Lett.}
  {\bfseries 91} (2003) 071101}
  [\href{https://arxiv.org/abs/astro-ph/0303596}{{\ttfamily
  astro-ph/0303596}}].

\bibitem{Lai:2022knd}
D.~Lai, \emph{{IXPE detection of polarized X-rays from magnetars and photon
  mode conversion at QED vacuum resonance}},
  \href{https://doi.org/10.1073/pnas.2216534120}{\emph{Proc. Nat. Acad. Sci.}
  {\bfseries 120} (2023) e2216534120}
  [\href{https://arxiv.org/abs/2209.13640}{{\ttfamily 2209.13640}}].

\bibitem{Heisenberg:1936nmg}
W.~Heisenberg and H.~Euler, \emph{{Consequences of Dirac's theory of
  positrons}}, \href{https://doi.org/10.1007/BF01343663}{\emph{Z. Phys.}
  {\bfseries 98} (1936) 714}
  [\href{https://arxiv.org/abs/physics/0605038}{{\ttfamily physics/0605038}}].

\bibitem{Schwinger:1951nm}
J.S.~Schwinger, \emph{{On gauge invariance and vacuum polarization}},
  \href{https://doi.org/10.1103/PhysRev.82.664}{\emph{Phys. Rev.} {\bfseries
  82} (1951) 664}.

\bibitem{Dunne:2004nc}
G.V.~Dunne, \emph{{Heisenberg-Euler effective Lagrangians: Basics and
  extensions}},  in \emph{{From fields to strings: Circumnavigating theoretical
  physics. Ian Kogan memorial collection (3 volume set)}}, M.~Shifman,
  A.~Vainshtein and J.~Wheater, eds., pp.~445--522 (2004),
  \href{https://doi.org/10.1142/9789812775344_0014}{DOI}
  [\href{https://arxiv.org/abs/hep-th/0406216}{{\ttfamily hep-th/0406216}}].

\bibitem{Ferrer:2012pb}
E.J.~Ferrer, V.~de~la Incera and A.~Sanchez, \emph{{Non-perturbative
  Euler-Heisenberg Lagrangian and Paraelectricity in Magnetized Massless QED}},
  \href{https://doi.org/10.1016/j.nuclphysb.2012.07.003}{\emph{Nucl. Phys. B}
  {\bfseries 864} (2012) 469}
  [\href{https://arxiv.org/abs/1204.3660}{{\ttfamily 1204.3660}}].

\bibitem{Baier:2007dw}
V.N.~Baier and V.M.~Katkov, \emph{{Pair creation by a photon in a strong
  magnetic field}},
  \href{https://doi.org/10.1103/PhysRevD.75.073009}{\emph{Phys. Rev. D}
  {\bfseries 75} (2007) 073009}
  [\href{https://arxiv.org/abs/hep-ph/0701119}{{\ttfamily hep-ph/0701119}}].

\bibitem{Hattori:2012je}
K.~Hattori and K.~Itakura, \emph{{Vacuum birefringence in strong magnetic
  fields: (I) Photon polarization tensor with all the Landau levels}},
  \href{https://doi.org/10.1016/j.aop.2012.11.010}{\emph{Annals Phys.}
  {\bfseries 330} (2013) 23} [\href{https://arxiv.org/abs/1209.2663}{{\ttfamily
  1209.2663}}].

\bibitem{Hattori:2012ny}
K.~Hattori and K.~Itakura, \emph{{Vacuum birefringence in strong magnetic
  fields: (II) Complex refractive index from the lowest Landau level}},
  \href{https://doi.org/10.1016/j.aop.2013.03.016}{\emph{Annals Phys.}
  {\bfseries 334} (2013) 58} [\href{https://arxiv.org/abs/1212.1897}{{\ttfamily
  1212.1897}}].

\bibitem{Alexandre:2000jc}
J.~Alexandre, \emph{{Vacuum polarization in thermal QED with an external
  magnetic field}},
  \href{https://doi.org/10.1103/PhysRevD.63.073010}{\emph{Phys. Rev. D}
  {\bfseries 63} (2001) 073010}
  [\href{https://arxiv.org/abs/hep-th/0009204}{{\ttfamily hep-th/0009204}}].

\bibitem{Wang:2020dsr}
X.~Wang, I.A.~Shovkovy, L.~Yu and M.~Huang, \emph{{Ellipticity of photon
  emission from strongly magnetized hot QCD plasma}},
  \href{https://doi.org/10.1103/PhysRevD.102.076010}{\emph{Phys. Rev. D}
  {\bfseries 102} (2020) 076010}
  [\href{https://arxiv.org/abs/2006.16254}{{\ttfamily 2006.16254}}].

\bibitem{Wang:2021ebh}
X.~Wang and I.~Shovkovy, \emph{{Photon polarization tensor in a magnetized
  plasma: Absorptive part}},
  \href{https://doi.org/10.1103/PhysRevD.104.056017}{\emph{Phys. Rev. D}
  {\bfseries 104} (2021) 056017}
  [\href{https://arxiv.org/abs/2103.01967}{{\ttfamily 2103.01967}}].

\bibitem{Wang:2021eud}
X.~Wang and I.~Shovkovy, \emph{{Polarization tensor of magnetized quark-gluon
  plasma at nonzero baryon density}},
  \href{https://doi.org/10.1140/epjc/s10052-021-09650-3}{\emph{Eur. Phys. J. C}
  {\bfseries 81} (2021) 901}
  [\href{https://arxiv.org/abs/2106.09029}{{\ttfamily 2106.09029}}].

\bibitem{Hattori:2022uzp}
K.~Hattori and K.~Itakura, \emph{{In-medium polarization tensor in strong
  magnetic fields (I): Magneto-birefringence at finite temperature and
  density}}, \href{https://doi.org/10.1016/j.aop.2022.169114}{\emph{Annals
  Phys.} {\bfseries 446} (2022) 169114}
  [\href{https://arxiv.org/abs/2205.04312}{{\ttfamily 2205.04312}}].

\bibitem{Hattori:2022wao}
K.~Hattori and K.~Itakura, \emph{{In-medium polarization tensor in strong
  magnetic fields (II): Axial Ward identity at finite temperature and
  density}}, \href{https://doi.org/10.1016/j.aop.2022.169115}{\emph{Annals
  Phys.} {\bfseries 446} (2022) 169115}
  [\href{https://arxiv.org/abs/2205.06411}{{\ttfamily 2205.06411}}].

\bibitem{Pyatkovskiy:2010xz}
P.K.~Pyatkovskiy and V.P.~Gusynin, \emph{{Dynamical polarization of monolayer
  graphene in a magnetic field}},
  \href{https://doi.org/10.1103/PhysRevB.83.075422}{\emph{Phys. Rev. B}
  {\bfseries 83} (2011) 075422}
  [\href{https://arxiv.org/abs/1009.5980}{{\ttfamily 1009.5980}}].

\bibitem{PerezRojas:1979jrk}
H.~Perez~Rojas and A.E.~Shabad, \emph{{POLARIZATION OF RELATIVISTIC ELECTRON
  AND POSITRON GAS IN A STRONG MAGNETIC FIELD. PROPAGATION OF ELECTROMAGNETIC
  WAVES}}, \href{https://doi.org/10.1016/0003-4916(79)90104-0}{\emph{Annals
  Phys.} {\bfseries 121} (1979) 432}.

\end{thebibliography}\endgroup

\end{document}